\documentclass[prb,aps,twocolumn,superscriptaddress,10pt,showpacs,longbibliography,noeprint]{revtex4-1}
\usepackage{graphicx}
\usepackage{epstopdf}
\usepackage{color}
\usepackage{amsmath,amssymb}
\usepackage{fixmath}

\newcommand{\angstrom}{\mbox{\normalfont\AA}}
\usepackage [english]{babel}
\usepackage [autostyle, english = american]{csquotes}
\MakeOuterQuote{"}
\usepackage{textgreek}
\usepackage[export]{adjustbox}
\usepackage{siunitx}
\usepackage{xcolor}
\usepackage{hyperref}
\usepackage{notes2bib}
\begin{document}
	
\title{Large dynamic scissoring mode displacements coupled to band gap opening in Hybrid Perovskites}

\author{Tobias A. Bird} 
\affiliation{Department of Chemistry, University of Warwick, Gibbet Hill, Coventry, CV4 7AL,United Kingdom}
\author{Jungshen Chen}
\affiliation{Nano-Science Center \& Department of Chemistry, University of Copenhagen, Universitetsparken 5, 2100 Copenhagen, Denmark}

\author{Manila Songvilay}
\affiliation{Institut N\'{e}el, CNRS and Universit\'{e} Grenoble Alpes, 38000 Grenoble, France}

\author{Chris Stock}
\affiliation{School of Physics and Astronomy, University of Edinburgh, Edinburgh EH9 3FD, United Kingdom}

\author{Michael T. Wharmby}
\affiliation{Deutsches Elektronen-Synchrotron (DESY), Notkestr. 85, 22607 Hamburg, Germany}

\author{Nicholas C. Bristowe}
\affiliation{Centre for Materials Physics, Durham University, South Road, Durham DH1 3LE, United Kingdom}

\author{Mark S. Senn}
\email{m.senn@warwick.ac.uk}
\affiliation{Department of Chemistry, University of Warwick, Gibbet Hill, Coventry, CV4 7AL,United Kingdom}

\date{\today}

\begin{abstract}
Hybrid perovskites are a rapidly growing research area, having reached photovoltaic power conversion efficiencies of over 25 \%. We apply a symmetry-motivated analysis method to analyse X-ray pair distribution function data of the cubic phases of the hybrid perovskites MAPb$X_3$ ($X$ = I, Br, Cl). We demonstrate that the local structure of the inorganic components of MAPb$X_3$ ($X$ = I, Br, Cl) are dominated by scissoring type deformations of the Pb$X_6$ octahedra. We find these modes to have a larger amplitude than equivalent distortions in the $A$-site deficient perovskite ScF$_3$ and demonstrate that they show a significant departure from the harmonic approximation. Calculations performed on an all-inorganic analogue to the hybrid perovskite, FrPbBr$_3$, show that the large amplitudes of the scissoring modes are coupled to an opening of the electronic band gap. Finally, we use density functional theory calculations to show that the organic MA cations reorientate to accomodate the large amplitude scissoring modes.
\end{abstract}

\maketitle

\section{Introduction}
Molecular perovskites, also known as hybrid perovskites, are a fast growing research area in photovoltaics, due to their low cost to make and rapid increase in efficiency (from 3.9 \% in 2009\cite{AkihiroKojimaKenjiroTeshimaYasuoShirai2009} to $>$ 25 \% today\cite{Green2014,Malinkiewicz2014,Schileo2020}). These materials have the general structure and chemical formula of traditional perovskites (AB$X_3$), but where they differ is that the A site cation is organic. The most frequently studied of this class of materials are the methylammonium (MA) lead halides, which have the general formula CH$_3$NH$_3$Pb$X_3$ ($X$ = I, Br, Cl), commonly abbreviated to MAPbX$_3$. In addition to their high conversion efficiency, this class of hybrid perovskites have other desirable photovoltaic properties, such as long charge carrier lifetimes\cite{Wehrenfennig2014}, mobility\cite{Zhao2017} and diffusion lengths\cite{Xing2013}, a high absorption coefficient\cite{Sun2014}, and a direct band gap\cite{AkihiroKojimaKenjiroTeshimaYasuoShirai2009}. These properties couple together to create a device that has a high density of charge carriers with a strong barrier against recombination, all whilst needing much less material than traditional solar cell materials, and without the need for a high energy input manufacturing process\cite{Hsiao2015}. 

Whilst perovskite oxides are a well studied class of materials due to the wide range of desirable properties exhibited by them, less is understood about the structure-property relationship in halide perovskites, particularly the hybrid perovskite family. Having a methylammonium ion rather than a metal ion at the A site results in the A site possessing an electric dipole moment rather than a point charge, so the dynamics of these ions are the focus of a lot of research in these hybrid perovskites. In the higher temperature tetragonal and cubic phases, the alignment of the ions appears to be disordered\cite{Leguy2016,Letoublon2016a,Brown2017,Songvilay2019}, however they could form small domains below the length scale required for coherent diffraction where the molecules are aligned. The dynamics of their rotations, and any local order, could have a large contribution to the properties of the material. For example, the interaction between phonons and the rotational degrees of freedom of the MA cations has been shown to have an impact on thermal conductivity\cite{Pisoni2014}. The changes in dynamics are thought to be closely linked to the structural changes of the material with temperature, and it is still unknown how the dynamics affect the properties of this material as a photovoltaic. Another question that has still not been fully solved is whether the configuration of the MA cations lead to this class of materials being ferroelectric\cite{Leguy2016,Hoque2016,Jankowska2017,Breternitz2020}.

Use of X-ray single crystal and powder diffraction has led to a good understanding of the different structural phases of these materials. Similarly to a large number of perovskites, all of the single-halide MAPbX$_3$ materials have cubic symmetry at high temperatures and undergo symmetry-lowering phase transitions to tetragonal and orthorhombic structures at lower temperatures\cite{Poglitsch1987}. Most experimental studies agree that there are 3 structural phases for MAPbI$_3$ and MAPbCl$_3$, however there is a 4$^{th}$ phase for MAPbBr$_3$ which is preferred for a small temperature range (\textit{ca.} 150-155 K), commonly thought to be an incommensurate phase\cite{Guo2017}. In the cubic phase, the MA cation is thought be fully disordered, with recent advances made using techniques such as NMR and quasi-elastic neutron scattering showing that the MA cation is close to having the orientational freedom of a free MA cation\cite{Chen2015,Leguy2015}. As the inorganic framework undergoes structural phase transitions, lowering the average symmetry from cubic $Pm\bar{3}m$, the orientational freedom of the MA cation is restricted, becoming fully ordered in the orthorhombic phases\cite{Weller2015,Brown2017}. This shows that the organic molecular and inorganic framework dynamics in MAPbX$_3$ are inherently linked\cite{Lee2015,Lee2016}. In addition to experimental studies, computational methods have seen a lot of use in this, and other, areas of research in hybrid perovskites\cite{Aristidou2017,Ghosh2017,Eames2015,Montero-Alejo2016,Carignano2015}. Both classical molecular dynamics and DFT simulations have demonstrated a link between the different phases of MAPbI$_3$ and the preferred orientations of the MA cations\cite{Mattoni2015a}. Work from Quarti \textit{et al} has demonstrated that the configuration of the MA cations has a significant effect on the properties of the material, such as its electronic band structure\cite{Quarti2014,Quarti2015}. This underlines why it is important to fully understand the structure-property relationship in these materials. Despite the knowledge that the organic molecular and inorganic framework dynamics are linked \textit{via} hydrogen bonding interactions, it is currently unclear how this interaction affects the dynamics as a whole.

The bands forming the top of the valence bands and the bottom of the conduction band in the electronic structure of the methylammonium lead halides will be dominated by Pb and $X$ ($X$ = I, Br, Cl) electrons\cite{Zhang2018,Lee2016}. Therefore, regardless of the role of the MA cation in stabilising particular distortions, it is necessary to establish good models for the dynamic distortions in the PbX$_3$ framework. In this work, we aim to probe the dynamics of the inorganic framework of the cubic phases of the three single-halide MAPbX$_3$ materials. We have recently demonstrated how by using a symmetry motivated approach to analysing PDF data we can gain extra information on disorder and dynamics within a system. Our study on BaTiO$_3$ has shown that this method is very sensitive to primary order parameters and is a powerful tool to analyse order-disorder phase transitions{\cite{Senn2016}}. Both this study and our more recent work on the negative thermal expansion materials ScF$_3$ and CaZrF$_6$ has demonstrated that this method is also sensitive to soft phonon modes and has also revealed substantial deviations from the crystallographic average structure in these materials\cite{Bird2020}. Here, we use X-ray total scattering data, which is much more sensitive to the inorganic framework than the molecular cations, to probe the characters of the low lying excitations of the cubic phases of the methylammonium lead halides.  

\section{Experimental Details and Data Analysis}
\begin{figure}
	\centering
	\includegraphics[width=\columnwidth]{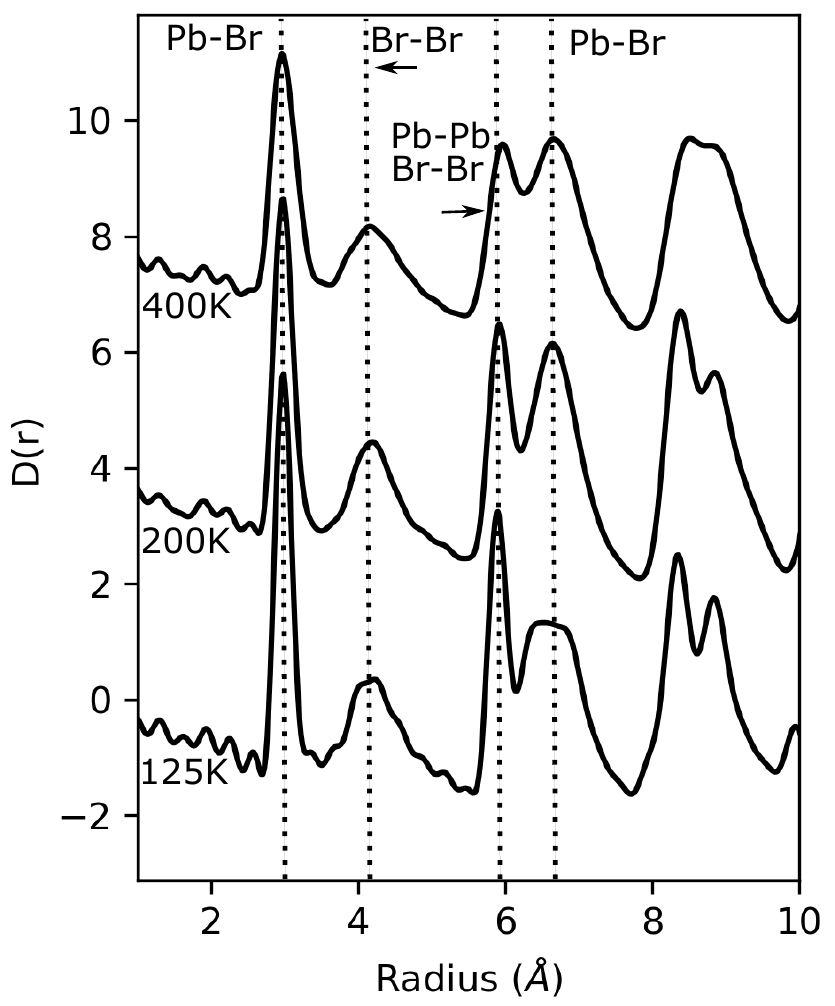}
	\caption{\label{pdfcomp}Pair distribution functions of MAPbBr$_3$ are shown in three different phases (orthorhomic, tetragonal and cubic, shown top to bottom). Each PDF is shown with a offset of 4 between them. Similar plots for X = I and Cl are given in the SI.}
\end{figure}
\begin{figure}[t!]
	\centering
	\includegraphics[width=\columnwidth]{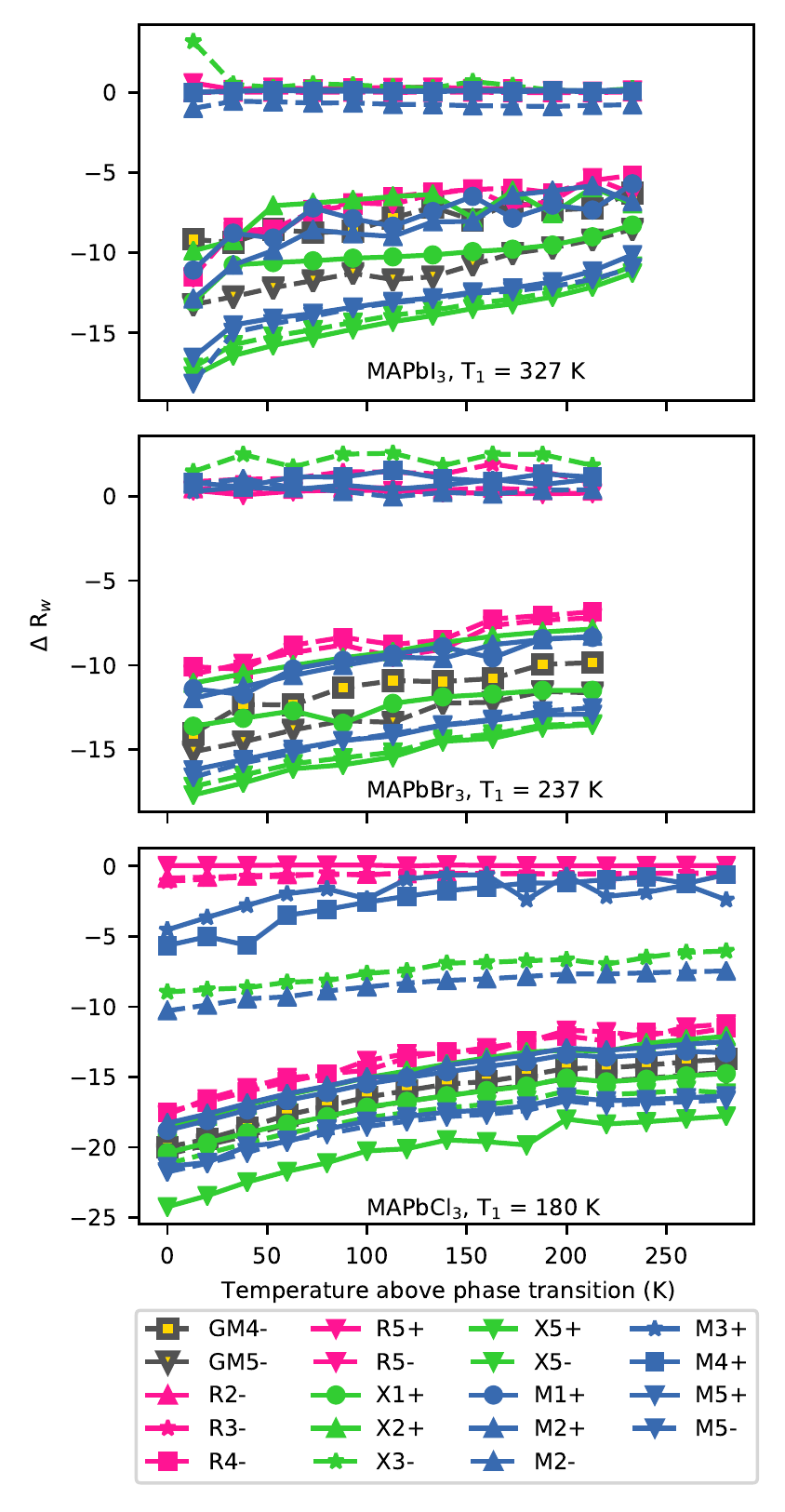}
	\caption{\label{f1} For each compound, the best individual fitting statistic is plotted for each irrep at each temperature. The R-factor is shown relative to the R-factor for the refinement with no symmetry adapted displacement modes active in the refinement, and the temperature shown is relative to the cubic phase transition as reported in the literature. The cubic transition temperature for each compound is indicated on the plot. The irreps are labelled as follows: colour denotes the k-point of the irrep, with blue referring to the M-point, green to X, pink to R and yellow to $\Gamma$; marker shape denotes the irrep number, with a circle referring to 1, an upward-pointed triangle to 2, a star to 3, a square to 4 and a downward-pointed triangle to 5; linestyle denotes the parity of the irrep, with a solid line referring to a "+" irrep, and a dashed line referring to a "-" irrep. }
\end{figure}

MAPbI$_3$ was prepared using the inverse temperature crystallisation method\cite{Saidaminov2015}. Briefly, equal molar amounts of MAI and PbI$_2$ were dissolved in a solvent (γ-butyrolactone) at room temperature. Then the obtained MAPbI$_3$ solution was heated to 110 $^{\circ}$C for the crystal growth. Powder samples of MAPbBr$_3$ were prepared by the reaction of stoichiometric amounts of lead acetate and methylamine hydrobromide in hydrobromic acid. The excess acid was then evaporated to leave an orange colored product which was washed with diethyl ether. Powder samples of MAPbCl$_3$ were prepared out of a solution of methylamine hydrochloride and lead acetate dissolved in hydrochloric acid. An excess of an approximately 8-10 molar ratio of methylamine hydrochloride was required to obtain these phase pure samples. The resulting powder was washed with diethyl ether.

For MAPbBr$_3$, synchrotron radiation X-ray total scattering experiments were conducted at the synchrotron facility PETRA III (beamline P02.1\cite{Dippel2015}) at DESY, Hamburg. A wavelength $\lambda = \SI{0.2070}{\angstrom} $ was used to collect data. Data were collected at temperatures of 125, 140, 147, 152 K and at intervals of 25 K from 175 to 450 K.

For MAPbI$_3$ and MAPbCl$_3$, Synchrotron radiation X-ray total scattering experiments were conducted at the synchrotron facility Diamond Light Source (beamline I15-1). A wavelength of $\lambda = \SI{0.161669}{\angstrom} $ was used to collect data. Data were collected at 20 K intervals over the temperature ranges 100 - 460 K (MAPbCl$_3$) and 100 - 560 K (MAPbI$_3$).

The obtained 2D images were masked and radially integrated using the DAWN\cite{Basham2015} software. $G(r)$ and $D(r)$ functions were computed using GudrunX\cite{McLain2012}, using $Q_{max}$ values of $21, 30$ and $\SI{28}{\angstrom}^{-1}$ for MAPbBr$_3$, MAPbCl$_3$ and MAPbI$_3$ respectively.  GudrunX was also used to perform background subtraction, sample absorption and fluorescence corrections. 

Analysis of the pair distribution functions was carried out using the symmetry-adapted PDF analysis (SAPA) method described in ref. 32. For each sample, a 2 $\times$ 2 $\times$ 2 $P1$ supercell of the $Pm\bar{3}m$ aristotype PbX$_3$ with Pb at (0.5, 0.5, 0.5) and $X$ at (0.5, 0.5, 0) was generated and parameterised in terms of symmetry adapted displacements using the ISODISTORT software\cite{Campbell2006}. The generated mode listings were output in .cif format and then converted to the .inp format of the TOPAS Academic software v6 using the Jedit macros\cite{Evans2010}. In total, there were 96 modes which transformed according to one of 19 irreducible representations. These supercells were generated without the organic A-site cation included, since the contribution of pairs involving the organic components of the structure will have a negligible contribution to the overall PDF due to their comparitively weak scattering power for X-rays. This lack of sensitivity of X-ray total scattering to the organic elements of hybrid perovskites can be seen by comparing recent publications by Malavasi \textit{et al}\cite{Page2016,Bernasconi2017,Bernasconi2018}. For each irreducible representation (irrep) at each temperature, refinements of the corresponding modes were started from random starting mode amplitudes. This was repeated 500 times. For all samples, the refinements were carried out with a fitting range of 1.7 to $\SI{20}{\angstrom}$. Refinements were also tested using a fitting range with a maximum of $\SI{10}{\angstrom}$ and found to be broadly similar.

The DFT calculations were performed using the Vienna Ab Initio Simulation Package (VASP)\cite{Kresse1994,Kresse1996,Kresse1996b,Kresse1993}, version 5.4.4. We employed the optB86b-vdW exchange correlation potential\cite{Klime2011} which includes VdW corrections previously found to suit hybrid perovskites\cite{Lee2016}. Projector augmented-wave (PAW) pseudopotentials\cite{Blochl1994,Kresse1996} were utilised, as supplied within the VASP package. A plane wave basis set with a 1100 eV energy cutoff and a 4 $\times$ 4 $\times$ 4 Monkhorst-Pack k-point mesh with respect to the parent cubic primitive cell (scaled accordingly for other supercells) were found suitable. The energy landscape of the various modes in the hybrid system were studied by fixing the halide framework while allowing for Pb and MA to relax until the forces were less than 5 meV/\si{\angstrom}. Results were compared with FrPbBr$_3$, which we used as a hypothetical inorganic analogue to the hybrid perovskite, since Fr best matches the ionic radii of MA\cite{Lee2016}.

\section{Results and Discussion}
\begin{figure*}[ht]
	\centering
	\includegraphics[]{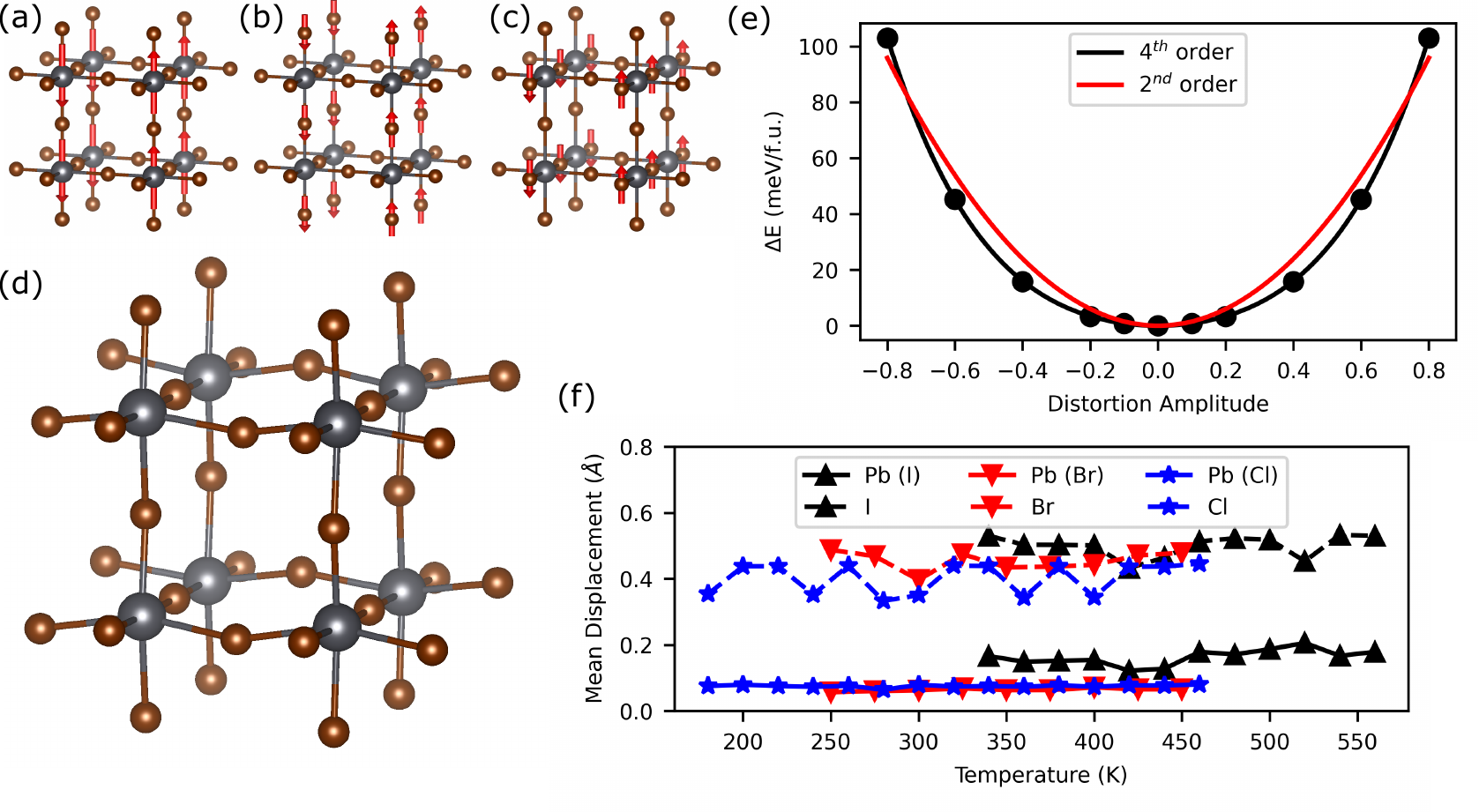}
	\caption{\label{x5p}(a-c) A breakdown of the atomic basis that spans the X$_5^+$ irrep. Shown in (d) is the structure resulting from a refinement of the $Pnma$ order parameter direction of the X$_5^+$ irrep. (e) Mode energies with varying distortion mode amplitude for the $Pnma$ order parameter direction of FrPbBr$_3$. Harmonic (2$^{nd}$ order) and anharmonic (4$^{th}$ order) fits to the potential well are shown. (f) Mean displacement values for the general X$_5^+$ order parameter direction. }
\end{figure*}

A key aspect of the local structure of the MAPb$X_3$ ($X$= I, Br, Cl) family of hybrid perovskites is that the first four peaks of the inorganic component of the PDF do not change much beyond that expected for simply changing the temperature, i.e a change in peak width corresponding to a change in thermal energy, and a change in peak position corresponding to thermal expansion. For MAPbBr$_3$ (Fig. \ref{pdfcomp}) and MAPbCl$_3$, the peaks stay the same from the low temperature orthorhombic phase into the high temperature cubic phase\cite{Bernasconi2017,Page2016,Bernasconi2018}. For MAPbI$_3$, there is a slight change upon the transition between the tetragonal and orthorhombic phases, but the peaks from the tetragonal phase persist in the cubic phase\cite{Beecher2016} (see SI). This has been taken to imply that the cubic phase consists of local symmetry-broken domains and there has been recent work to support this hypothesis\cite{Beecher2016}. This would suggest that the distortions most responsible for the local structure should be the rigid-unit modes (RUMs) that drive these phase transitions. 

To gain a more robust understanding of the local structure of MAPb$X_3$ ($X$= I, Br, Cl), we perform a symmetry-adapted PDF analysis (SAPA)\cite{Bird2020,Senn2016} to eludicate the character of the dominant lattice dynamics associated with the inorganic cage. We note that we are insensitive to MA orientation and displacement modes in the present X-ray PDF study, and so no attempt is made to model these against the experimental data. The symmetry-adapted displacements which show the most improvement in the $R_w$ for the models against the PDFs for all compounds and all temperatures are those which transform according to irreps that permit a scissoring motion of the $X$ anions, \textit{i.e.}, the Br--Pb--Br bond angles are distorted away from 90$^{\circ}$ but the Pb--Br bond lengths remain undistorted. This result of the SAPA analysis does not imply that the RUMs are high energy modes, it simply means that the majority of the motion of the halide anions arise from these scissoring modes. This is supported by competetive two-phase refinements of the PDFs, in which we allow the X$_5^+$ displacements to refine in one phase and the displacements for one of the RUMs (R$_5^-$ or M$_2^+$) in the other. These refinements show a preference for scissoring modes compared to the RUMs for all 3 samples, as evident from the refined scale factors of the two phases which show an approximate scissoring:rotation ratio of 2.3:1 (see SI for more details). For context, this ratio is approximately 4:1 in ScF$_3$, which is isostructural to the inorganic framework of MAPb$X_3$. The lower ratio compared to ScF$_3$ reflects a lower flexibility due to the presence of an $A$-site cation, which can interact with the inorganic framework \textit{via} hydrogen bonding\cite{Lee2015}. However, it is clear from our results that the majority of the halide anion motion still arises from scissoring-type deformations of the octahedra.

The above results are in line with a recent reverse Monte Carlo (RMC) analysis of neutron PDFs of MAPbI$_3$\cite{Liu2017} between 10 and 400 K. This study demonstrates that a bending of the Pb-I-Pb bond angle dominates the local distortions of the PbI$_6$ octahedra. Our results show that the four best fitting modes all have scissoring character, of which it is the X$_5^+$ (Fig \ref{x5p} (c\&d)) that performs best across all three compositions and temperatures. This could be due to the fact that there are more parameters for the X$_5^+$ irrep than the other three (18 modes transform as the X$_5^+$ irrep, compared to 12 for M$_5^-$ and 6 for X$_5^-$ and M$_5^+$), but the improvement could arise from the anti-polar Pb displacements that enter into the irrep X$_5^+$, although this is unlikely since they only have a small contribution to the overall displacements. The three distortions that span this irrep are shown in Fig \ref{x5p}. We find the amplitudes of these scissoring modes to be quite large; refinements of X$_5^+$ and X$_5^-$ in the tetragonal phase of MAPbBr$_3$ resulted in supercell-normalised mode amplitudes of $\approx$ $\SI{1.35}{\angstrom}$. This is close in magnitude to the equivalent amplitude of the R$_5^-$ distortion ($\approx$ $\SI{1.65}{\angstrom}$) which is frozen into the structure in the tetragonal phase.

Given how large the local deviations are from the average structure, it is reasonable to assume they will have a substantial effect on the band structure. We used DFT calculations to investigate the impact that the scissoring modes could have on the electronic band structure of the hybrid perovskites. We chose to analyse MAPbBr$_3$, since it is cubic at room temperature where experimental band gap values have been reported, and to focus on the two X point modes that do the best job at describing the deviations away from local cubic symmetry, as evident in the PDF data. For a completely unrestrained order parameter direction transforming as X$_5^+$, there are a rather large number of degrees of freedom (18 in total), so, to make our results more robust, and to facilitate a direct comparison to X$_5^-$, we take results from refinements using higher symmetry OPDs with no more than 5 parameters. We use structures from refinements against our data with X$_5^+$ OPDs with $Pnma$ and $Cmcm$ symmetry ($(0,a;b,0;0,c)$ and $(0,a;b,b;a,0)$ respectively) and the X$_5^-$ OPD with $C2/c$ symmetry ($(a,b;c,-c;-b,-a)$) as input to our band structure calculations. For the two X$_5^+$ OPDs, only Br anion displacements were refined when generating the CIFs for the band structure calculations, although by symmetry, Pb displacements also enter into the irrep. For X$_5^-$, Pb displacements are forbidden by symmetry. We also sampled points of different overall distortion amplitude along the X$_5^+$ OPD with $Pnma$ symmetry and calculated the energy. These energy calculations were performed for the FrPbBr$_3$ structures used to calculate the band structure.

\begin{figure}[t]
	\centering
	\includegraphics[width=\columnwidth]{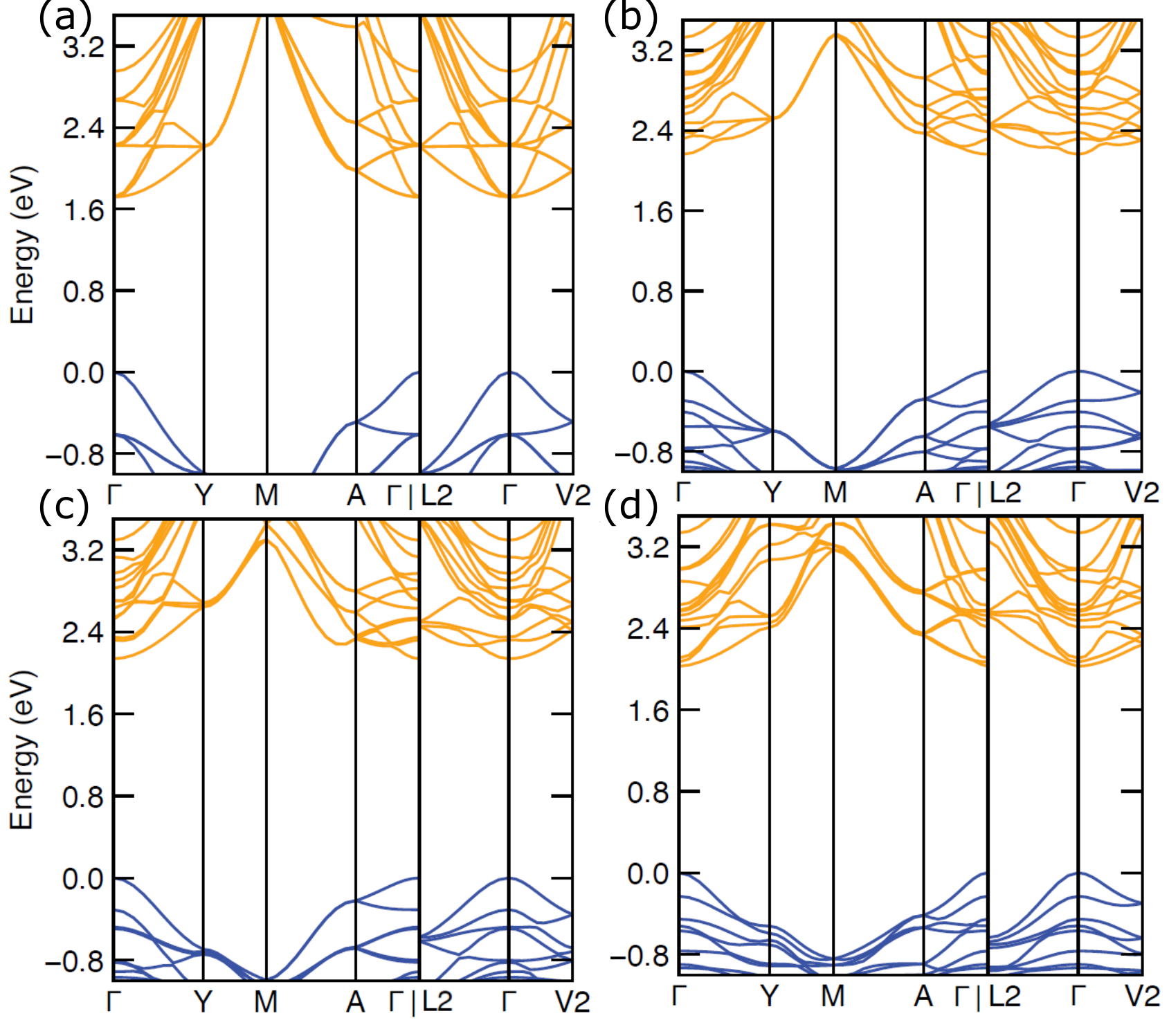}
	\caption{\label{bands}The calculated electronic band structure of FrPbBr$_3$ for the undistorted structure (a), the $Pnma$ and $Cmcm$ order parameter directions of the X$_5^+$ irrep (b and c, respectively) and the $C2/c$ order parameter direction of the X5$_5^-$ irrep (d). These figures were created using sumo\cite{MGanose2018}}
\end{figure}

In the undistorted structure, the calculated band gap was 1.717 eV, which is slightly higher than other calculated band gaps for cubic MAPbBr$_3$ (1.64 eV\cite{Mosconi2016}) at the same level of theory, and is direct. Previous work has shown that substitution of Fr for MA opens up the band gap slightly in orthorhombic MAPbI$_3$\cite{Lee2016}. For each distortion, the band gap opens up significantly to values of 2.025, 2.138 and 2.162 eV for the $C2/c$, $Cmcm$ and $Pnma$ distortions with an amplitude of 0.8$\times$ the maximum amplitude refined from PDF data for the X$_5^+$ distortions and 1.1$\times$ the maximum amplitude for X$_5^-$, respectively, and remains direct. These relative amplitudes were chosen so all 3 distortions were at similar mode amplitudes.  These values are closer to the experimentally determined band gaps for MAPbBr$_3$ of $\approx$ 2.3 eV at room temperature\cite{Papavassiliou1995}, although this is likely due to a cancellation of errors. The distortions result in a reduced orbital overlap between Pb and Br p-orbitals, leading to a lower band curvature and therefore an increased effective mass in the distorted band structures (Fig \ref{bands} and SI). The mobility of polarons is inversely proportional to the electron band effective mass\cite{Frost2017}, and this increased effective mass in the distorted structures may explain the discrepancy between experimental and calculated values\cite{Bonn2017}.

Spin-orbit coupling (SOC) interactions, which play a large role in systems involving Pb, have not been accounted for. Consequently, the exact shape of the electron bands and size of the band gap won't be accurate, since inclusion of SOC has been shown to lead to unconventional dispersion relations\cite{Brivio2014}. The effects from SOC on band gap size in halide perovskites tend to be canceled out by full treatment of electron Coulomb interactions beyond DFT\cite{Even2013,Umari2014}. Therefore, the trends we detect due to the different distortion modes will remain the same. In the two X$_5^+$ distortions, the degeneracy of the bands at the conduction band minimum (CBM) at the \textGamma point are broken, leading to fewer available states at the CBM. Contrastingly, the X$_5^+$ $Pnma$ distortion appears to have the largest DOS at the valence band maximum due to the reduced bandwidth. Fluctuations in the band gap of hybrid perovskites due to their highly dynamic structure has been previously predicted\cite{Quarti2015}, and is expected to assist the initial stages of charge separation. In addition, an increase of the band gap coinciding with a transverse displacement of I ions in MAPbI$_3$ due to an external strain field has been reported\cite{Zhang2018}.

Our refinements against the PDF data show that all three modes have a large amplitude, with supercell-normalised mode amplitudes of 1.84, 1.82 and $\SI{1.36}{\angstrom}$ for OPDs with $Pnma$, $Cmcm$ and $C2/c$ symmetries, respectively. These mode amplitudes correspond to maxiumum Br displacements of 0.486, 0.410 and $\SI{0.350}{\angstrom}$. Note that the refined distortions correspond to a time-averaged view of the structure, so these maximum Br displacements are a factor of $\sqrt{2} \times$ greater, in the harmonic approximation, than those found in the refined structures. As a consequence of their large amplitudes, the distortions would be expected to be anharmonic in nature, which is supported by the potential energy well we calculate for the X$_5^+$ $(0,a;b,0;0,c)$ OPD in FrPbBr$_3$ (Fig \ref{x5p} (f)), which has a significant quartic component when fit with a 4$^{th}$ order polynomial fit ($\Delta E = 127 x^4 + 79.7 x^2$, where $x$ is the distortion mode amplitude relative to its maximum value at 400 K). This breakdown in the harmonic approximation would then allow the scissoring modes of the inorganic framework to couple directly to the anharmonic modes that correspond to the organic cation dynamics\cite{Li2017,Ghosh2017a}. Despite the presence of an A-site in these materials, the amplitude of these scissoring modes are greater than those in ScF$_3$, suggesting the MA cations move to accomodate the large-amplitude modes. The implication of this, then, is that the band gap opening we detect as a response to the scissoring modes is likely influenced by the dynamics of the MA cations, although our refinements are only sensitive to the inorganic framework.

To investigate the above hypothesis, we consider the X$_5^+$ OPD with $Pnma$ symmetry. This breaks the equivalency of the $< 1 0 0 >$ directions and leads to two distinct A-site symmetries (see SI). Therefore, if the inorganic and organic dynamics are coupled together, we would expect to see the MA cations located at different points of the unit cell to respond differently to the distortion mode, to reflect the different local environments they would experience. To test this, we relaxed the MA cations from an initial anti-polar configuration with the C-N bonds aligned with the [1 0 0] direction, in a structure with a 0.8$\times$ X$_5^+$ $(0,a;b,0;0,c)$ distortion (relative to the maximum amplitude at 400 K) frozen in. The MA cations showed significant reorientation, with the "edge" ((0.5, 0, 0) and equivalents) and "corner" ((0, 0, 0)) cations rotating to include significant components along $c$. There is a split amongst the "face" cations, with two (at (0.5, 0.5, 0) and (0, 0.5, 0.5)) rotating to include smaller components along the $b$- and $c$-axes. The remaining "face" cation and the cation located at the centre of the supercell both rotate to include a significant component along $c$ and a smaller component along $b$. In all, there are 5 distinct C--N bond alignments, which may reflect the 5 distinct Br sites. In addition, all cations show a slight displacement from the high-symmetry-unique positions. Full details can be found in the SI. This demonstrates that the MA cations can rotate to accommodate the distortions of the inorganic framework, indicating that the dynamics of the two components of the structure may be linked. However, it is important to note that our calculations are effectively performed at 0 K, where the ground state is the fully ordered orthorhombic phase. It is quite possible that the configurational entropy associated with the MA orderings may effectively act to decouple these dynamics at higher temperatures in the cubic phase. Indeed, there is evidence to suggest the organic and inorganic dynamics are decoupled in MAPbCl$_3$\cite{Songvilay2019}. Additionally, a similar computational result in CsPbBr$_3$ showing coupling between large amplitude distortions of the Br ions and head-to-head Cs motion\cite{Yaffe2017} suggest this feature may not be exclusive to hybrid inorganic systems. We have also shown that acoustic phonon lifetimes for the all-inorganic CsPbBr$_3$ are very similar to those in MAPbCl$_3$\cite{Songvilay2019a}, further supporting the idea that at high temperatures the MA rotational modes may have little effect on the lattice phonon modes.

There has been recent literature support for the idea that cubic halide perovskites, rather than being treated as a single repeating unit, should be thought of as a network of polymorphs showing different symmetry-lowering deformations of the average structure, such as varying degrees of octahedral tilting or differing amplitudes of B-site displacement\cite{Zhao2020}. Our work is broadly consistent with this picture. The sterochemical behaviour of the Pb cation, in conjunction with the coupling between organic cation and inorganic framework dynamics, is likely to have a large impact on the possible polymorphs the material exhibits within this hypothesis of the nature of the structure of halide perovskites.

In summary, we have shown that large scissoring modes of the halide ions describe the dominant deviations from the average structure in the cubic phases of the hybrid perovskites. These modes have a similar amplitude to those of the static RUMs below the phase transition temperature. These distortions have the effect of opening up the band gap of the electronic structure. In addition, we have shown that the organic cations can move to accomodate the distortions of the inorganic framework, suggesting the dynamics of the two components could be inherently linked, and that the inorganic lattice is likely to be significantly distorted from the average at a local level. These dynamic structures should be accounted for in simulations performed on the hybrid perovskites, since they can have a significant effect on the calculated properties.

The PDF data used in this study are available as a supporting \href{https://figshare.com/articles/dataset/Data_for_Large_dynamic_scissoring_mode_displacements_coupled_to_band_gap_opening_in_hybrid_perovskites_/15155658}{dataset}.

\section*{Acknowledgements}
T.A.B thanks EPSRC for a PhD studentship through the EPSRC Centre for Doctoral Training in Molecular Analytical Science, grant number EP/L015307/1. M.S.S acknowledges the Royal Society for a University Research Fellowship (UF160265). N.C.B acknowledges computational resources from the Hamilton HPC Service of Durham University and the UK Materials and Molecular Modelling Hub (partially funded by the EPSRC project EP/P020194/1). We acknowledge DESY (Hamburg, Germany), a member of the Helmholtz Association HGF, for the provision of experimental facilities. Parts of this research were carried out at PETRA III. We also thank Diamond Light Source for providing experiment time on beamline I15-1 under proposal number CY21611.  Samples were characterised via Block Allocation Group Award (EE18786) at the high resolution powder diffractometer I11, Diamond Light Source.

\bibliography{mapbx3_refs.bib}

\begin{thebibliography}{65}%
\makeatletter
\providecommand \@ifxundefined [1]{%
 \@ifx{#1\undefined}
}%
\providecommand \@ifnum [1]{%
 \ifnum #1\expandafter \@firstoftwo
 \else \expandafter \@secondoftwo
 \fi
}%
\providecommand \@ifx [1]{%
 \ifx #1\expandafter \@firstoftwo
 \else \expandafter \@secondoftwo
 \fi
}%
\providecommand \natexlab [1]{#1}%
\providecommand \enquote  [1]{``#1''}%
\providecommand \bibnamefont  [1]{#1}%
\providecommand \bibfnamefont [1]{#1}%
\providecommand \citenamefont [1]{#1}%
\providecommand \href@noop [0]{\@secondoftwo}%
\providecommand \href [0]{\begingroup \@sanitize@url \@href}%
\providecommand \@href[1]{\@@startlink{#1}\@@href}%
\providecommand \@@href[1]{\endgroup#1\@@endlink}%
\providecommand \@sanitize@url [0]{\catcode `\\12\catcode `\$12\catcode
  `\&12\catcode `\#12\catcode `\^12\catcode `\_12\catcode `\%12\relax}%
\providecommand \@@startlink[1]{}%
\providecommand \@@endlink[0]{}%
\providecommand \url  [0]{\begingroup\@sanitize@url \@url }%
\providecommand \@url [1]{\endgroup\@href {#1}{\urlprefix }}%
\providecommand \urlprefix  [0]{URL }%
\providecommand \Eprint [0]{\href }%
\providecommand \doibase [0]{http://dx.doi.org/}%
\providecommand \selectlanguage [0]{\@gobble}%
\providecommand \bibinfo  [0]{\@secondoftwo}%
\providecommand \bibfield  [0]{\@secondoftwo}%
\providecommand \translation [1]{[#1]}%
\providecommand \BibitemOpen [0]{}%
\providecommand \bibitemStop [0]{}%
\providecommand \bibitemNoStop [0]{.\EOS\space}%
\providecommand \EOS [0]{\spacefactor3000\relax}%
\providecommand \BibitemShut  [1]{\csname bibitem#1\endcsname}%
\let\auto@bib@innerbib\@empty
\bibitem [{\citenamefont {Kojima}\ \emph {et~al.}(2009)\citenamefont {Kojima},
  \citenamefont {Teshima}, \citenamefont {Shirai},\ and\ \citenamefont
  {Miyasaka}}]{AkihiroKojimaKenjiroTeshimaYasuoShirai2009}%
  \BibitemOpen
  \bibfield  {author} {\bibinfo {author} {\bibfnamefont {Akihiro}\ \bibnamefont
  {Kojima}}, \bibinfo {author} {\bibfnamefont {Kenjiro}\ \bibnamefont
  {Teshima}}, \bibinfo {author} {\bibfnamefont {Yasuo}\ \bibnamefont {Shirai}},
  \ and\ \bibinfo {author} {\bibfnamefont {Tsutomu}\ \bibnamefont {Miyasaka}},\
  }\bibfield  {title} {\enquote {\bibinfo {title} {{Organometal Halide
  Perovskites as Visible- Light Sensitizers for Photovoltaic Cells}},}\ }\href
  {\doibase 10.1021/ja809598r} {\bibfield  {journal} {\bibinfo  {journal} {J.
  Am. Chem. Soc.}\ }\textbf {\bibinfo {volume} {131}},\ \bibinfo {pages}
  {6050--6051} (\bibinfo {year} {2009})}\BibitemShut {NoStop}%
\bibitem [{\citenamefont {Green}\ \emph {et~al.}(2014)\citenamefont {Green},
  \citenamefont {Ho-Baillie},\ and\ \citenamefont {Snaith}}]{Green2014}%
  \BibitemOpen
  \bibfield  {author} {\bibinfo {author} {\bibfnamefont {Martin~A.}\
  \bibnamefont {Green}}, \bibinfo {author} {\bibfnamefont {Anita}\ \bibnamefont
  {Ho-Baillie}}, \ and\ \bibinfo {author} {\bibfnamefont {Henry~J.}\
  \bibnamefont {Snaith}},\ }\bibfield  {title} {\enquote {\bibinfo {title}
  {{The emergence of perovskite solar cells}},}\ }\href {\doibase
  10.1038/nphoton.2014.134} {\bibfield  {journal} {\bibinfo  {journal} {Nat.
  Photonics}\ }\textbf {\bibinfo {volume} {8}},\ \bibinfo {pages} {506--514}
  (\bibinfo {year} {2014})}\BibitemShut {NoStop}%
\bibitem [{\citenamefont {Malinkiewicz}\ \emph {et~al.}(2014)\citenamefont
  {Malinkiewicz}, \citenamefont {Yella}, \citenamefont {Lee}, \citenamefont
  {Espallargas}, \citenamefont {Graetzel}, \citenamefont {Nazeeruddin},\ and\
  \citenamefont {Bolink}}]{Malinkiewicz2014}%
  \BibitemOpen
  \bibfield  {author} {\bibinfo {author} {\bibfnamefont {Olga}\ \bibnamefont
  {Malinkiewicz}}, \bibinfo {author} {\bibfnamefont {Aswani}\ \bibnamefont
  {Yella}}, \bibinfo {author} {\bibfnamefont {Yong~Hui}\ \bibnamefont {Lee}},
  \bibinfo {author} {\bibfnamefont {Guillermo~M{\'{i}}nguez}\ \bibnamefont
  {Espallargas}}, \bibinfo {author} {\bibfnamefont {Michael}\ \bibnamefont
  {Graetzel}}, \bibinfo {author} {\bibfnamefont {Mohammad~K.}\ \bibnamefont
  {Nazeeruddin}}, \ and\ \bibinfo {author} {\bibfnamefont {Henk~J.}\
  \bibnamefont {Bolink}},\ }\bibfield  {title} {\enquote {\bibinfo {title}
  {{Perovskite solar cells employing organic charge-transport layers}},}\
  }\href {\doibase 10.1038/nphoton.2013.341} {\bibfield  {journal} {\bibinfo
  {journal} {Nat. Photonics}\ }\textbf {\bibinfo {volume} {8}},\ \bibinfo
  {pages} {128--132} (\bibinfo {year} {2014})}\BibitemShut {NoStop}%
\bibitem [{\citenamefont {Schileo}\ and\ \citenamefont
  {Grancini}(2020)}]{Schileo2020}%
  \BibitemOpen
  \bibfield  {author} {\bibinfo {author} {\bibfnamefont {Giorgio}\ \bibnamefont
  {Schileo}}\ and\ \bibinfo {author} {\bibfnamefont {Giulia}\ \bibnamefont
  {Grancini}},\ }\bibfield  {title} {\enquote {\bibinfo {title} {{Halide
  perovskites: current issues and new strategies to push material and device
  stability}},}\ }\href {\doibase 10.1088/2515-7655/ab6cc4} {\bibfield
  {journal} {\bibinfo  {journal} {J. Phys.: Energy}\ }\textbf {\bibinfo
  {volume} {2}},\ \bibinfo {pages} {021005} (\bibinfo {year}
  {2020})}\BibitemShut {NoStop}%
\bibitem [{\citenamefont {Wehrenfennig}\ \emph {et~al.}(2014)\citenamefont
  {Wehrenfennig}, \citenamefont {Eperon}, \citenamefont {Johnston},
  \citenamefont {Snaith},\ and\ \citenamefont {Herz}}]{Wehrenfennig2014}%
  \BibitemOpen
  \bibfield  {author} {\bibinfo {author} {\bibfnamefont {Christian}\
  \bibnamefont {Wehrenfennig}}, \bibinfo {author} {\bibfnamefont {Giles~E.}\
  \bibnamefont {Eperon}}, \bibinfo {author} {\bibfnamefont {Michael~B.}\
  \bibnamefont {Johnston}}, \bibinfo {author} {\bibfnamefont {Henry~J.}\
  \bibnamefont {Snaith}}, \ and\ \bibinfo {author} {\bibfnamefont {Laura~M.}\
  \bibnamefont {Herz}},\ }\bibfield  {title} {\enquote {\bibinfo {title} {{High
  charge carrier mobilities and lifetimes in organolead trihalide
  perovskites}},}\ }\href {\doibase 10.1002/adma.201305172} {\bibfield
  {journal} {\bibinfo  {journal} {Adv. Mater.}\ }\textbf {\bibinfo {volume}
  {26}},\ \bibinfo {pages} {1584--1589} (\bibinfo {year} {2014})}\BibitemShut
  {NoStop}%
\bibitem [{\citenamefont {Zhao}\ \emph {et~al.}(2017)\citenamefont {Zhao},
  \citenamefont {Skelton}, \citenamefont {Hu}, \citenamefont {La-o vorakiat},
  \citenamefont {Zhu}, \citenamefont {Marcus}, \citenamefont {Michel-Beyerle},
  \citenamefont {Lam}, \citenamefont {Walsh},\ and\ \citenamefont
  {Chia}}]{Zhao2017}%
  \BibitemOpen
  \bibfield  {author} {\bibinfo {author} {\bibfnamefont {Daming}\ \bibnamefont
  {Zhao}}, \bibinfo {author} {\bibfnamefont {Jonathan~M.}\ \bibnamefont
  {Skelton}}, \bibinfo {author} {\bibfnamefont {Hongwei}\ \bibnamefont {Hu}},
  \bibinfo {author} {\bibfnamefont {Chan}\ \bibnamefont {La-o vorakiat}},
  \bibinfo {author} {\bibfnamefont {Jian-Xin}\ \bibnamefont {Zhu}}, \bibinfo
  {author} {\bibfnamefont {Rudolph~A.}\ \bibnamefont {Marcus}}, \bibinfo
  {author} {\bibfnamefont {Maria-Elisabeth}\ \bibnamefont {Michel-Beyerle}},
  \bibinfo {author} {\bibfnamefont {Yeng~Ming}\ \bibnamefont {Lam}}, \bibinfo
  {author} {\bibfnamefont {Aron}\ \bibnamefont {Walsh}}, \ and\ \bibinfo
  {author} {\bibfnamefont {Elbert E.~M.}\ \bibnamefont {Chia}},\ }\bibfield
  {title} {\enquote {\bibinfo {title} {{Low-frequency optical phonon modes and
  carrier mobility in the halide perovskite CH 3 NH 3 PbBr 3 using terahertz
  time-domain spectroscopy}},}\ }\href {\doibase 10.1063/1.4993524} {\bibfield
  {journal} {\bibinfo  {journal} {Appl. Phys. Lett.}\ }\textbf {\bibinfo
  {volume} {111}},\ \bibinfo {pages} {201903} (\bibinfo {year}
  {2017})}\BibitemShut {NoStop}%
\bibitem [{\citenamefont {Xing}\ \emph {et~al.}(2013)\citenamefont {Xing},
  \citenamefont {Mathews}, \citenamefont {Sun}, \citenamefont {Lim},
  \citenamefont {Lam}, \citenamefont {Gr{\"{a}}tzel}, \citenamefont
  {Mhaisalkar},\ and\ \citenamefont {Sum}}]{Xing2013}%
  \BibitemOpen
  \bibfield  {author} {\bibinfo {author} {\bibfnamefont {Guichuan}\
  \bibnamefont {Xing}}, \bibinfo {author} {\bibfnamefont {Nripan}\ \bibnamefont
  {Mathews}}, \bibinfo {author} {\bibfnamefont {Shuangyong}\ \bibnamefont
  {Sun}}, \bibinfo {author} {\bibfnamefont {Swee~Sien}\ \bibnamefont {Lim}},
  \bibinfo {author} {\bibfnamefont {Yeng~Ming}\ \bibnamefont {Lam}}, \bibinfo
  {author} {\bibfnamefont {Michael}\ \bibnamefont {Gr{\"{a}}tzel}}, \bibinfo
  {author} {\bibfnamefont {Subodh}\ \bibnamefont {Mhaisalkar}}, \ and\ \bibinfo
  {author} {\bibfnamefont {Tze~Chien}\ \bibnamefont {Sum}},\ }\bibfield
  {title} {\enquote {\bibinfo {title} {{Long-range balanced electron- and
  hole-transport lengths in organic-inorganic CH3NH3PbI3.}}}\ }\href {\doibase
  10.1126/science.1243167} {\bibfield  {journal} {\bibinfo  {journal}
  {Science}\ }\textbf {\bibinfo {volume} {342}},\ \bibinfo {pages} {344--347}
  (\bibinfo {year} {2013})}\BibitemShut {NoStop}%
\bibitem [{\citenamefont {Sun}\ \emph {et~al.}(2014)\citenamefont {Sun},
  \citenamefont {Salim}, \citenamefont {Mathews}, \citenamefont {Duchamp},
  \citenamefont {Boothroyd}, \citenamefont {Xing}, \citenamefont {Sum},\ and\
  \citenamefont {Lam}}]{Sun2014}%
  \BibitemOpen
  \bibfield  {author} {\bibinfo {author} {\bibfnamefont {Shuangyong}\
  \bibnamefont {Sun}}, \bibinfo {author} {\bibfnamefont {Teddy}\ \bibnamefont
  {Salim}}, \bibinfo {author} {\bibfnamefont {Nripan}\ \bibnamefont {Mathews}},
  \bibinfo {author} {\bibfnamefont {Martial}\ \bibnamefont {Duchamp}}, \bibinfo
  {author} {\bibfnamefont {Chris}\ \bibnamefont {Boothroyd}}, \bibinfo {author}
  {\bibfnamefont {Guichuan}\ \bibnamefont {Xing}}, \bibinfo {author}
  {\bibfnamefont {Tze~Chien}\ \bibnamefont {Sum}}, \ and\ \bibinfo {author}
  {\bibfnamefont {Yeng~Ming}\ \bibnamefont {Lam}},\ }\bibfield  {title}
  {\enquote {\bibinfo {title} {{The origin of high efficiency in
  low-temperature solution-processable bilayer organometal halide hybrid solar
  cells}},}\ }\href {\doibase 10.1039/c3ee43161d} {\bibfield  {journal}
  {\bibinfo  {journal} {Energy Environ. Sci.}\ }\textbf {\bibinfo {volume}
  {7}},\ \bibinfo {pages} {399--407} (\bibinfo {year} {2014})}\BibitemShut
  {NoStop}%
\bibitem [{\citenamefont {Hsiao}\ \emph {et~al.}(2015)\citenamefont {Hsiao},
  \citenamefont {Wu}, \citenamefont {Li}, \citenamefont {Liu}, \citenamefont
  {Qin},\ and\ \citenamefont {Hu}}]{Hsiao2015}%
  \BibitemOpen
  \bibfield  {author} {\bibinfo {author} {\bibfnamefont {Yu-Che}\ \bibnamefont
  {Hsiao}}, \bibinfo {author} {\bibfnamefont {Ting}\ \bibnamefont {Wu}},
  \bibinfo {author} {\bibfnamefont {Mingxing}\ \bibnamefont {Li}}, \bibinfo
  {author} {\bibfnamefont {Qing}\ \bibnamefont {Liu}}, \bibinfo {author}
  {\bibfnamefont {Wei}\ \bibnamefont {Qin}}, \ and\ \bibinfo {author}
  {\bibfnamefont {Bin}\ \bibnamefont {Hu}},\ }\bibfield  {title} {\enquote
  {\bibinfo {title} {{Fundamental physics behind high-efficiency organo-metal
  halide perovskite solar cells}},}\ }\href {\doibase 10.1039/C5TA01376C}
  {\bibfield  {journal} {\bibinfo  {journal} {J. Mater. Chem. A}\ }\textbf
  {\bibinfo {volume} {3}},\ \bibinfo {pages} {15372--15385} (\bibinfo {year}
  {2015})}\BibitemShut {NoStop}%
\bibitem [{\citenamefont {Leguy}\ \emph {et~al.}(2016)\citenamefont {Leguy},
  \citenamefont {Go{\~{n}}i}, \citenamefont {Frost}, \citenamefont {Skelton},
  \citenamefont {Brivio}, \citenamefont {Rodr{\'{i}}guez-Mart{\'{i}}nez},
  \citenamefont {Weber}, \citenamefont {Pallipurath}, \citenamefont {Alonso},
  \citenamefont {Campoy-Quiles}, \citenamefont {Weller}, \citenamefont
  {Nelson}, \citenamefont {Walsh},\ and\ \citenamefont {Barnes}}]{Leguy2016}%
  \BibitemOpen
  \bibfield  {author} {\bibinfo {author} {\bibfnamefont {Aur{\'{e}}lien M.~A.}\
  \bibnamefont {Leguy}}, \bibinfo {author} {\bibfnamefont {Alejandro~R.}\
  \bibnamefont {Go{\~{n}}i}}, \bibinfo {author} {\bibfnamefont {Jarvist~M.}\
  \bibnamefont {Frost}}, \bibinfo {author} {\bibfnamefont {Jonathan}\
  \bibnamefont {Skelton}}, \bibinfo {author} {\bibfnamefont {Federico}\
  \bibnamefont {Brivio}}, \bibinfo {author} {\bibfnamefont {Xabier}\
  \bibnamefont {Rodr{\'{i}}guez-Mart{\'{i}}nez}}, \bibinfo {author}
  {\bibfnamefont {Oliver~J.}\ \bibnamefont {Weber}}, \bibinfo {author}
  {\bibfnamefont {Anuradha}\ \bibnamefont {Pallipurath}}, \bibinfo {author}
  {\bibfnamefont {M.~Isabel}\ \bibnamefont {Alonso}}, \bibinfo {author}
  {\bibfnamefont {Mariano}\ \bibnamefont {Campoy-Quiles}}, \bibinfo {author}
  {\bibfnamefont {Mark~T.}\ \bibnamefont {Weller}}, \bibinfo {author}
  {\bibfnamefont {Jenny}\ \bibnamefont {Nelson}}, \bibinfo {author}
  {\bibfnamefont {Aron}\ \bibnamefont {Walsh}}, \ and\ \bibinfo {author}
  {\bibfnamefont {Piers R.~F.}\ \bibnamefont {Barnes}},\ }\bibfield  {title}
  {\enquote {\bibinfo {title} {{Dynamic disorder, phonon lifetimes, and the
  assignment of modes to the vibrational spectra of methylammonium lead halide
  perovskites}},}\ }\href {\doibase 10.1039/C6CP03474H} {\bibfield  {journal}
  {\bibinfo  {journal} {Phys. Chem. Chem. Phys.}\ }\textbf {\bibinfo {volume}
  {18}},\ \bibinfo {pages} {27051--27066} (\bibinfo {year} {2016})}\BibitemShut
  {NoStop}%
\bibitem [{\citenamefont {L{\'{e}}toublon}\ \emph {et~al.}(2016)\citenamefont
  {L{\'{e}}toublon}, \citenamefont {Paofai}, \citenamefont {Ruffl{\'{e}}},
  \citenamefont {Bourges}, \citenamefont {Hehlen}, \citenamefont {Michel},
  \citenamefont {Ecolivet}, \citenamefont {Durand}, \citenamefont {Cordier},
  \citenamefont {Katan},\ and\ \citenamefont {Even}}]{Letoublon2016a}%
  \BibitemOpen
  \bibfield  {author} {\bibinfo {author} {\bibfnamefont {Antoine}\ \bibnamefont
  {L{\'{e}}toublon}}, \bibinfo {author} {\bibfnamefont {Serge}\ \bibnamefont
  {Paofai}}, \bibinfo {author} {\bibfnamefont {Beno{\^{i}}t}\ \bibnamefont
  {Ruffl{\'{e}}}}, \bibinfo {author} {\bibfnamefont {Philippe}\ \bibnamefont
  {Bourges}}, \bibinfo {author} {\bibfnamefont {Bernard}\ \bibnamefont
  {Hehlen}}, \bibinfo {author} {\bibfnamefont {Thierry}\ \bibnamefont
  {Michel}}, \bibinfo {author} {\bibfnamefont {Claude}\ \bibnamefont
  {Ecolivet}}, \bibinfo {author} {\bibfnamefont {Olivier}\ \bibnamefont
  {Durand}}, \bibinfo {author} {\bibfnamefont {St{\'{e}}phane}\ \bibnamefont
  {Cordier}}, \bibinfo {author} {\bibfnamefont {Claudine}\ \bibnamefont
  {Katan}}, \ and\ \bibinfo {author} {\bibfnamefont {Jacky}\ \bibnamefont
  {Even}},\ }\bibfield  {title} {\enquote {\bibinfo {title} {{Elastic
  Constants, Optical Phonons, and Molecular Relaxations in the High Temperature
  Plastic Phase of the CH3NH3PbBr3 Hybrid Perovskite}},}\ }\href {\doibase
  10.1021/acs.jpclett.6b01709} {\bibfield  {journal} {\bibinfo  {journal} {J.
  Phys. Chem. Lett.}\ }\textbf {\bibinfo {volume} {7}},\ \bibinfo {pages}
  {3776--3784} (\bibinfo {year} {2016})}\BibitemShut {NoStop}%
\bibitem [{\citenamefont {Brown}\ \emph {et~al.}(2017)\citenamefont {Brown},
  \citenamefont {Parker}, \citenamefont {Garc{\'{i}}a}, \citenamefont
  {Mukhopadhyay}, \citenamefont {Sakai},\ and\ \citenamefont
  {Stock}}]{Brown2017}%
  \BibitemOpen
  \bibfield  {author} {\bibinfo {author} {\bibfnamefont {K.~L.}\ \bibnamefont
  {Brown}}, \bibinfo {author} {\bibfnamefont {S.~F.}\ \bibnamefont {Parker}},
  \bibinfo {author} {\bibfnamefont {I.~Robles}\ \bibnamefont {Garc{\'{i}}a}},
  \bibinfo {author} {\bibfnamefont {S.}~\bibnamefont {Mukhopadhyay}}, \bibinfo
  {author} {\bibfnamefont {V.~Garc{\'{i}}a}\ \bibnamefont {Sakai}}, \ and\
  \bibinfo {author} {\bibfnamefont {C.}~\bibnamefont {Stock}},\ }\bibfield
  {title} {\enquote {\bibinfo {title} {{Molecular orientational melting within
  a lead-halide octahedron framework: The order-disorder transition in
  CH3NH3PbBr3}},}\ }\href {\doibase 10.1103/PhysRevB.96.174111} {\bibfield
  {journal} {\bibinfo  {journal} {Phys. Rev. B}\ }\textbf {\bibinfo {volume}
  {96}},\ \bibinfo {pages} {174111} (\bibinfo {year} {2017})}\BibitemShut
  {NoStop}%
\bibitem [{\citenamefont {Songvilay}\ \emph
  {et~al.}(2019{\natexlab{a}})\citenamefont {Songvilay}, \citenamefont {Wang},
  \citenamefont {Sakai}, \citenamefont {Guidi}, \citenamefont {Bari},
  \citenamefont {Ye}, \citenamefont {Xu}, \citenamefont {Brown}, \citenamefont
  {Gehring},\ and\ \citenamefont {Stock}}]{Songvilay2019}%
  \BibitemOpen
  \bibfield  {author} {\bibinfo {author} {\bibfnamefont {M.}~\bibnamefont
  {Songvilay}}, \bibinfo {author} {\bibfnamefont {Zitian}\ \bibnamefont
  {Wang}}, \bibinfo {author} {\bibfnamefont {V.~Garcia}\ \bibnamefont {Sakai}},
  \bibinfo {author} {\bibfnamefont {T.}~\bibnamefont {Guidi}}, \bibinfo
  {author} {\bibfnamefont {M.}~\bibnamefont {Bari}}, \bibinfo {author}
  {\bibfnamefont {Z.~G.}\ \bibnamefont {Ye}}, \bibinfo {author} {\bibfnamefont
  {Guangyong}\ \bibnamefont {Xu}}, \bibinfo {author} {\bibfnamefont {K.~L.}\
  \bibnamefont {Brown}}, \bibinfo {author} {\bibfnamefont {P.~M.}\ \bibnamefont
  {Gehring}}, \ and\ \bibinfo {author} {\bibfnamefont {C.}~\bibnamefont
  {Stock}},\ }\bibfield  {title} {\enquote {\bibinfo {title} {{Decoupled
  molecular and inorganic framework dynamics in CH3NH3PbCl3}},}\ }\href
  {\doibase 10.1103/PhysRevMaterials.3.125406} {\bibfield  {journal} {\bibinfo
  {journal} {Phys. Rev. Mater.}\ }\textbf {\bibinfo {volume} {3}},\ \bibinfo
  {pages} {125406} (\bibinfo {year} {2019}{\natexlab{a}})}\BibitemShut
  {NoStop}%
\bibitem [{\citenamefont {Pisoni}\ \emph {et~al.}(2014)\citenamefont {Pisoni},
  \citenamefont {Ja{\'{c}}imovi{\'{c}}}, \citenamefont {Bari{\v{s}}i{\'{c}}},
  \citenamefont {Spina}, \citenamefont {Ga{\'{a}}l}, \citenamefont
  {Forr{\'{o}}},\ and\ \citenamefont {Horv{\'{a}}th}}]{Pisoni2014}%
  \BibitemOpen
  \bibfield  {author} {\bibinfo {author} {\bibfnamefont {Andrea}\ \bibnamefont
  {Pisoni}}, \bibinfo {author} {\bibfnamefont {Ja{\'{c}}im}\ \bibnamefont
  {Ja{\'{c}}imovi{\'{c}}}}, \bibinfo {author} {\bibfnamefont {Osor~S.}\
  \bibnamefont {Bari{\v{s}}i{\'{c}}}}, \bibinfo {author} {\bibfnamefont
  {Massimo}\ \bibnamefont {Spina}}, \bibinfo {author} {\bibfnamefont {Richard}\
  \bibnamefont {Ga{\'{a}}l}}, \bibinfo {author} {\bibfnamefont
  {L{\'{a}}szl{\'{o}}}\ \bibnamefont {Forr{\'{o}}}}, \ and\ \bibinfo {author}
  {\bibfnamefont {Endre}\ \bibnamefont {Horv{\'{a}}th}},\ }\bibfield  {title}
  {\enquote {\bibinfo {title} {{Ultra-low thermal conductivity in
  organic-inorganic hybrid perovskite CH3NH3PbI3}},}\ }\href {\doibase
  10.1021/jz5012109} {\bibfield  {journal} {\bibinfo  {journal} {J. Phys. Chem.
  Lett.}\ }\textbf {\bibinfo {volume} {5}},\ \bibinfo {pages} {2488--2492}
  (\bibinfo {year} {2014})}\BibitemShut {NoStop}%
\bibitem [{\citenamefont {Hoque}\ \emph {et~al.}(2016)\citenamefont {Hoque},
  \citenamefont {Yang}, \citenamefont {Li}, \citenamefont {Islam},
  \citenamefont {Pan}, \citenamefont {Zhu},\ and\ \citenamefont
  {Fan}}]{Hoque2016}%
  \BibitemOpen
  \bibfield  {author} {\bibinfo {author} {\bibfnamefont {M~Nadim~Ferdous}\
  \bibnamefont {Hoque}}, \bibinfo {author} {\bibfnamefont {Mengjin}\
  \bibnamefont {Yang}}, \bibinfo {author} {\bibfnamefont {Zhen}\ \bibnamefont
  {Li}}, \bibinfo {author} {\bibfnamefont {Nazifah}\ \bibnamefont {Islam}},
  \bibinfo {author} {\bibfnamefont {Xuan}\ \bibnamefont {Pan}}, \bibinfo
  {author} {\bibfnamefont {Kai}\ \bibnamefont {Zhu}}, \ and\ \bibinfo {author}
  {\bibfnamefont {Zhaoyang}\ \bibnamefont {Fan}},\ }\bibfield  {title}
  {\enquote {\bibinfo {title} {{Polarization and Dielectric Study of
  Methylammonium Lead Iodide Thin Film to Reveal its Nonferroelectric Nature
  under Solar Cell Operating Conditions}},}\ }\href {\doibase
  10.1021/acsenergylett.6b00093} {\bibfield  {journal} {\bibinfo  {journal}
  {ACS Energy Lett.}\ }\textbf {\bibinfo {volume} {1}},\ \bibinfo {pages}
  {142--149} (\bibinfo {year} {2016})}\BibitemShut {NoStop}%
\bibitem [{\citenamefont {Jankowska}\ and\ \citenamefont
  {Prezhdo}(2017)}]{Jankowska2017}%
  \BibitemOpen
  \bibfield  {author} {\bibinfo {author} {\bibfnamefont {Joanna}\ \bibnamefont
  {Jankowska}}\ and\ \bibinfo {author} {\bibfnamefont {Oleg~V.}\ \bibnamefont
  {Prezhdo}},\ }\bibfield  {title} {\enquote {\bibinfo {title} {{Ferroelectric
  Alignment of Organic Cations Inhibits Nonradiative Electron-Hole
  Recombination in Hybrid Perovskites: Ab Initio Nonadiabatic Molecular
  Dynamics}},}\ }\href {\doibase 10.1021/acs.jpclett.7b00008} {\bibfield
  {journal} {\bibinfo  {journal} {J. Phys. Chem. Lett.}\ }\textbf {\bibinfo
  {volume} {8}},\ \bibinfo {pages} {812--818} (\bibinfo {year}
  {2017})}\BibitemShut {NoStop}%
\bibitem [{\citenamefont {Breternitz}\ \emph {et~al.}(2020)\citenamefont
  {Breternitz}, \citenamefont {Lehmann}, \citenamefont {Barnett}, \citenamefont
  {Nowell},\ and\ \citenamefont {Schorr}}]{Breternitz2020}%
  \BibitemOpen
  \bibfield  {author} {\bibinfo {author} {\bibfnamefont {J.}~\bibnamefont
  {Breternitz}}, \bibinfo {author} {\bibfnamefont {F.}~\bibnamefont {Lehmann}},
  \bibinfo {author} {\bibfnamefont {S.~A.}\ \bibnamefont {Barnett}}, \bibinfo
  {author} {\bibfnamefont {H.}~\bibnamefont {Nowell}}, \ and\ \bibinfo {author}
  {\bibfnamefont {S.}~\bibnamefont {Schorr}},\ }\bibfield  {title} {\enquote
  {\bibinfo {title} {{Role of the Iodide–Methylammonium Interaction in the
  Ferroelectricity of CH3NH3PbI3}},}\ }\href {\doibase 10.1002/anie.201910599}
  {\bibfield  {journal} {\bibinfo  {journal} {Angew. Chem. Int. Ed.}\ }\textbf
  {\bibinfo {volume} {59}},\ \bibinfo {pages} {424--428} (\bibinfo {year}
  {2020})}\BibitemShut {NoStop}%
\bibitem [{\citenamefont {Poglitsch}\ and\ \citenamefont
  {Weber}(1987)}]{Poglitsch1987}%
  \BibitemOpen
  \bibfield  {author} {\bibinfo {author} {\bibfnamefont {A.}~\bibnamefont
  {Poglitsch}}\ and\ \bibinfo {author} {\bibfnamefont {D.}~\bibnamefont
  {Weber}},\ }\bibfield  {title} {\enquote {\bibinfo {title} {{Dynamic disorder
  in methylammoniumtrihalogenoplumbates (II) observed by millimeter-wave
  spectroscopy}},}\ }\href {\doibase 10.1063/1.453467} {\bibfield  {journal}
  {\bibinfo  {journal} {J. Chem. Phys.}\ }\textbf {\bibinfo {volume} {87}},\
  \bibinfo {pages} {6373--6378} (\bibinfo {year} {1987})}\BibitemShut {NoStop}%
\bibitem [{\citenamefont {Guo}\ \emph {et~al.}(2017)\citenamefont {Guo},
  \citenamefont {Yaffe}, \citenamefont {Paley}, \citenamefont {Beecher},
  \citenamefont {Hull}, \citenamefont {Szpak}, \citenamefont {Owen},
  \citenamefont {Brus},\ and\ \citenamefont {Pimenta}}]{Guo2017}%
  \BibitemOpen
  \bibfield  {author} {\bibinfo {author} {\bibfnamefont {Yinsheng}\
  \bibnamefont {Guo}}, \bibinfo {author} {\bibfnamefont {Omer}\ \bibnamefont
  {Yaffe}}, \bibinfo {author} {\bibfnamefont {Daniel~W.}\ \bibnamefont
  {Paley}}, \bibinfo {author} {\bibfnamefont {Alexander~N.}\ \bibnamefont
  {Beecher}}, \bibinfo {author} {\bibfnamefont {Trevor~D.}\ \bibnamefont
  {Hull}}, \bibinfo {author} {\bibfnamefont {Guilherme}\ \bibnamefont {Szpak}},
  \bibinfo {author} {\bibfnamefont {Jonathan~S.}\ \bibnamefont {Owen}},
  \bibinfo {author} {\bibfnamefont {Louis~E.}\ \bibnamefont {Brus}}, \ and\
  \bibinfo {author} {\bibfnamefont {Marcos~A.}\ \bibnamefont {Pimenta}},\
  }\bibfield  {title} {\enquote {\bibinfo {title} {{Interplay between organic
  cations and inorganic framework and incommensurability in hybrid lead-halide
  perovskite CH3NH3PbBr3}},}\ }\href {\doibase
  10.1103/PhysRevMaterials.1.042401} {\bibfield  {journal} {\bibinfo  {journal}
  {Phys. Rev. Mater.}\ }\textbf {\bibinfo {volume} {1}},\ \bibinfo {pages}
  {042401} (\bibinfo {year} {2017})}\BibitemShut {NoStop}%
\bibitem [{\citenamefont {Chen}\ \emph {et~al.}(2015)\citenamefont {Chen},
  \citenamefont {Foley}, \citenamefont {Ipek}, \citenamefont {Tyagi},
  \citenamefont {Copley}, \citenamefont {Brown}, \citenamefont {Choi},\ and\
  \citenamefont {Lee}}]{Chen2015}%
  \BibitemOpen
  \bibfield  {author} {\bibinfo {author} {\bibfnamefont {Tianran}\ \bibnamefont
  {Chen}}, \bibinfo {author} {\bibfnamefont {Benjamin~J.}\ \bibnamefont
  {Foley}}, \bibinfo {author} {\bibfnamefont {Bahar}\ \bibnamefont {Ipek}},
  \bibinfo {author} {\bibfnamefont {Madhusudan}\ \bibnamefont {Tyagi}},
  \bibinfo {author} {\bibfnamefont {John R.~D.}\ \bibnamefont {Copley}},
  \bibinfo {author} {\bibfnamefont {Craig~M.}\ \bibnamefont {Brown}}, \bibinfo
  {author} {\bibfnamefont {Joshua~J.}\ \bibnamefont {Choi}}, \ and\ \bibinfo
  {author} {\bibfnamefont {Seung-Hun}\ \bibnamefont {Lee}},\ }\bibfield
  {title} {\enquote {\bibinfo {title} {{Rotational dynamics of organic cations
  in the CH$_3$NH$_3$PbI$_3$ perovskite}},}\ }\href {\doibase
  10.1039/C5CP05348J} {\bibfield  {journal} {\bibinfo  {journal} {Phys. Chem.
  Chem. Phys.}\ }\textbf {\bibinfo {volume} {17}},\ \bibinfo {pages}
  {31278--31286} (\bibinfo {year} {2015})}\BibitemShut {NoStop}%
\bibitem [{\citenamefont {Leguy}\ \emph {et~al.}(2015)\citenamefont {Leguy},
  \citenamefont {Frost}, \citenamefont {McMahon}, \citenamefont {Sakai},
  \citenamefont {Kochelmann}, \citenamefont {Law}, \citenamefont {Li},
  \citenamefont {Foglia}, \citenamefont {Walsh}, \citenamefont {O'Regan},
  \citenamefont {Nelson}, \citenamefont {Cabral},\ and\ \citenamefont
  {Barnes}}]{Leguy2015}%
  \BibitemOpen
  \bibfield  {author} {\bibinfo {author} {\bibfnamefont {Aurelien~M.A.}\
  \bibnamefont {Leguy}}, \bibinfo {author} {\bibfnamefont {Jarvist~Moore}\
  \bibnamefont {Frost}}, \bibinfo {author} {\bibfnamefont {Andrew~P.}\
  \bibnamefont {McMahon}}, \bibinfo {author} {\bibfnamefont {Victoria~Garcia}\
  \bibnamefont {Sakai}}, \bibinfo {author} {\bibfnamefont {W.}~\bibnamefont
  {Kochelmann}}, \bibinfo {author} {\bibfnamefont {Chunhung}\ \bibnamefont
  {Law}}, \bibinfo {author} {\bibfnamefont {Xiaoe}\ \bibnamefont {Li}},
  \bibinfo {author} {\bibfnamefont {Fabrizia}\ \bibnamefont {Foglia}}, \bibinfo
  {author} {\bibfnamefont {Aron}\ \bibnamefont {Walsh}}, \bibinfo {author}
  {\bibfnamefont {Brian~C.}\ \bibnamefont {O'Regan}}, \bibinfo {author}
  {\bibfnamefont {Jenny}\ \bibnamefont {Nelson}}, \bibinfo {author}
  {\bibfnamefont {Jo{\~{a}}o~T.}\ \bibnamefont {Cabral}}, \ and\ \bibinfo
  {author} {\bibfnamefont {Piers~R.F.}\ \bibnamefont {Barnes}},\ }\bibfield
  {title} {\enquote {\bibinfo {title} {{The dynamics of methylammonium ions in
  hybrid organic-inorganic perovskite solar cells}},}\ }\href {\doibase
  10.1038/ncomms8124} {\bibfield  {journal} {\bibinfo  {journal} {Nat.
  Commun.}\ }\textbf {\bibinfo {volume} {6}},\ \bibinfo {pages} {7124}
  (\bibinfo {year} {2015})}\BibitemShut {NoStop}%
\bibitem [{\citenamefont {Weller}\ \emph {et~al.}(2015)\citenamefont {Weller},
  \citenamefont {Weber}, \citenamefont {Henry}, \citenamefont {{Di Pumpo}},\
  and\ \citenamefont {Hansen}}]{Weller2015}%
  \BibitemOpen
  \bibfield  {author} {\bibinfo {author} {\bibfnamefont {Mark~T.}\ \bibnamefont
  {Weller}}, \bibinfo {author} {\bibfnamefont {Oliver~J.}\ \bibnamefont
  {Weber}}, \bibinfo {author} {\bibfnamefont {Paul~F.}\ \bibnamefont {Henry}},
  \bibinfo {author} {\bibfnamefont {Antonietta~M.}\ \bibnamefont {{Di Pumpo}}},
  \ and\ \bibinfo {author} {\bibfnamefont {Thomas~C.}\ \bibnamefont {Hansen}},\
  }\bibfield  {title} {\enquote {\bibinfo {title} {{Complete structure and
  cation orientation in the perovskite photovoltaic methylammonium lead iodide
  between 100 and 352 K}},}\ }\href {\doibase 10.1039/c4cc09944c} {\bibfield
  {journal} {\bibinfo  {journal} {Chem. Commun.}\ }\textbf {\bibinfo {volume}
  {51}},\ \bibinfo {pages} {4180--4183} (\bibinfo {year} {2015})}\BibitemShut
  {NoStop}%
\bibitem [{\citenamefont {Lee}\ \emph {et~al.}(2015)\citenamefont {Lee},
  \citenamefont {Bristowe}, \citenamefont {Bristowe},\ and\ \citenamefont
  {Cheetham}}]{Lee2015}%
  \BibitemOpen
  \bibfield  {author} {\bibinfo {author} {\bibfnamefont {Jung~Hoon}\
  \bibnamefont {Lee}}, \bibinfo {author} {\bibfnamefont {Nicholas~C.}\
  \bibnamefont {Bristowe}}, \bibinfo {author} {\bibfnamefont {Paul~D.}\
  \bibnamefont {Bristowe}}, \ and\ \bibinfo {author} {\bibfnamefont
  {Anthony~K.}\ \bibnamefont {Cheetham}},\ }\bibfield  {title} {\enquote
  {\bibinfo {title} {{Role of hydrogen-bonding and its interplay with
  octahedral tilting in CH3NH3PbI3}},}\ }\href {\doibase 10.1039/c5cc00979k}
  {\bibfield  {journal} {\bibinfo  {journal} {Chem. Commun.}\ }\textbf
  {\bibinfo {volume} {51}},\ \bibinfo {pages} {6434--6437} (\bibinfo {year}
  {2015})}\BibitemShut {NoStop}%
\bibitem [{\citenamefont {Lee}\ \emph {et~al.}(2016)\citenamefont {Lee},
  \citenamefont {Bristowe}, \citenamefont {Lee}, \citenamefont {Lee},
  \citenamefont {Bristowe}, \citenamefont {Cheetham},\ and\ \citenamefont
  {Jang}}]{Lee2016}%
  \BibitemOpen
  \bibfield  {author} {\bibinfo {author} {\bibfnamefont {Jung~Hoon}\
  \bibnamefont {Lee}}, \bibinfo {author} {\bibfnamefont {Nicholas~C.}\
  \bibnamefont {Bristowe}}, \bibinfo {author} {\bibfnamefont {June~Ho}\
  \bibnamefont {Lee}}, \bibinfo {author} {\bibfnamefont {Sung~Hoon}\
  \bibnamefont {Lee}}, \bibinfo {author} {\bibfnamefont {Paul~D.}\ \bibnamefont
  {Bristowe}}, \bibinfo {author} {\bibfnamefont {Anthony~K.}\ \bibnamefont
  {Cheetham}}, \ and\ \bibinfo {author} {\bibfnamefont {Hyun~Myung}\
  \bibnamefont {Jang}},\ }\bibfield  {title} {\enquote {\bibinfo {title}
  {{Resolving the Physical Origin of Octahedral Tilting in Halide
  Perovskites}},}\ }\href {\doibase 10.1021/acs.chemmater.6b00968} {\bibfield
  {journal} {\bibinfo  {journal} {Chem. Mater.}\ }\textbf {\bibinfo {volume}
  {28}},\ \bibinfo {pages} {4259--4266} (\bibinfo {year} {2016})}\BibitemShut
  {NoStop}%
\bibitem [{\citenamefont {Aristidou}\ \emph {et~al.}(2017)\citenamefont
  {Aristidou}, \citenamefont {Eames}, \citenamefont {Sanchez-Molina},
  \citenamefont {Bu}, \citenamefont {Kosco}, \citenamefont {{Saiful Islam}},\
  and\ \citenamefont {Haque}}]{Aristidou2017}%
  \BibitemOpen
  \bibfield  {author} {\bibinfo {author} {\bibfnamefont {Nicholas}\
  \bibnamefont {Aristidou}}, \bibinfo {author} {\bibfnamefont {Christopher}\
  \bibnamefont {Eames}}, \bibinfo {author} {\bibfnamefont {Irene}\ \bibnamefont
  {Sanchez-Molina}}, \bibinfo {author} {\bibfnamefont {Xiangnan}\ \bibnamefont
  {Bu}}, \bibinfo {author} {\bibfnamefont {Jan}\ \bibnamefont {Kosco}},
  \bibinfo {author} {\bibfnamefont {M.}~\bibnamefont {{Saiful Islam}}}, \ and\
  \bibinfo {author} {\bibfnamefont {Saif~A.}\ \bibnamefont {Haque}},\
  }\bibfield  {title} {\enquote {\bibinfo {title} {{Fast oxygen diffusion and
  iodide defects mediate oxygen-induced degradation of perovskite solar
  cells}},}\ }\href {\doibase 10.1038/ncomms15218} {\bibfield  {journal}
  {\bibinfo  {journal} {Nat. Commun.}\ }\textbf {\bibinfo {volume} {8}},\
  \bibinfo {pages} {15218} (\bibinfo {year} {2017})}\BibitemShut {NoStop}%
\bibitem [{\citenamefont {Ghosh}\ \emph
  {et~al.}(2017{\natexlab{a}})\citenamefont {Ghosh}, \citenamefont {{Walsh
  Atkins}}, \citenamefont {Islam}, \citenamefont {Walker},\ and\ \citenamefont
  {Eames}}]{Ghosh2017}%
  \BibitemOpen
  \bibfield  {author} {\bibinfo {author} {\bibfnamefont {Dibyajyoti}\
  \bibnamefont {Ghosh}}, \bibinfo {author} {\bibfnamefont {Philip}\
  \bibnamefont {{Walsh Atkins}}}, \bibinfo {author} {\bibfnamefont {M.~Saiful}\
  \bibnamefont {Islam}}, \bibinfo {author} {\bibfnamefont {Alison~B.}\
  \bibnamefont {Walker}}, \ and\ \bibinfo {author} {\bibfnamefont
  {Christopher}\ \bibnamefont {Eames}},\ }\bibfield  {title} {\enquote
  {\bibinfo {title} {{Good Vibrations: Locking of Octahedral Tilting in
  Mixed-Cation Iodide Perovskites for Solar Cells}},}\ }\href {\doibase
  10.1021/acsenergylett.7b00729} {\bibfield  {journal} {\bibinfo  {journal}
  {ACS Energy Lett.}\ }\textbf {\bibinfo {volume} {2}},\ \bibinfo {pages}
  {2424--2429} (\bibinfo {year} {2017}{\natexlab{a}})}\BibitemShut {NoStop}%
\bibitem [{\citenamefont {Eames}\ \emph {et~al.}(2015)\citenamefont {Eames},
  \citenamefont {Frost}, \citenamefont {Barnes}, \citenamefont {O'Regan},
  \citenamefont {Walsh},\ and\ \citenamefont {Islam}}]{Eames2015}%
  \BibitemOpen
  \bibfield  {author} {\bibinfo {author} {\bibfnamefont {Christopher}\
  \bibnamefont {Eames}}, \bibinfo {author} {\bibfnamefont {Jarvist~M.}\
  \bibnamefont {Frost}}, \bibinfo {author} {\bibfnamefont {Piers~R.F.}\
  \bibnamefont {Barnes}}, \bibinfo {author} {\bibfnamefont {Brian~C.}\
  \bibnamefont {O'Regan}}, \bibinfo {author} {\bibfnamefont {Aron}\
  \bibnamefont {Walsh}}, \ and\ \bibinfo {author} {\bibfnamefont {M.~Saiful}\
  \bibnamefont {Islam}},\ }\bibfield  {title} {\enquote {\bibinfo {title}
  {{Ionic transport in hybrid lead iodide perovskite solar cells}},}\ }\href
  {\doibase 10.1038/ncomms8497} {\bibfield  {journal} {\bibinfo  {journal}
  {Nat. Commun.}\ }\textbf {\bibinfo {volume} {6}},\ \bibinfo {pages} {7497}
  (\bibinfo {year} {2015})}\BibitemShut {NoStop}%
\bibitem [{\citenamefont {Montero-Alejo}\ \emph {et~al.}(2016)\citenamefont
  {Montero-Alejo}, \citenamefont {Men{\'{e}}ndez-Proupin}, \citenamefont
  {Hidalgo-Rojas}, \citenamefont {Palacios}, \citenamefont {Wahn{\'{o}}n},\
  and\ \citenamefont {Conesa}}]{Montero-Alejo2016}%
  \BibitemOpen
  \bibfield  {author} {\bibinfo {author} {\bibfnamefont {Ana~L.}\ \bibnamefont
  {Montero-Alejo}}, \bibinfo {author} {\bibfnamefont {E.}~\bibnamefont
  {Men{\'{e}}ndez-Proupin}}, \bibinfo {author} {\bibfnamefont {D.}~\bibnamefont
  {Hidalgo-Rojas}}, \bibinfo {author} {\bibfnamefont {P.}~\bibnamefont
  {Palacios}}, \bibinfo {author} {\bibfnamefont {P.}~\bibnamefont
  {Wahn{\'{o}}n}}, \ and\ \bibinfo {author} {\bibfnamefont {J.~C.}\
  \bibnamefont {Conesa}},\ }\bibfield  {title} {\enquote {\bibinfo {title}
  {{Modeling of Thermal Effect on the Electronic Properties of Photovoltaic
  Perovskite CH3NH3PbI3: The Case of Tetragonal Phase}},}\ }\href {\doibase
  10.1021/acs.jpcc.6b01013} {\bibfield  {journal} {\bibinfo  {journal} {J.
  Phys. Chem. C}\ }\textbf {\bibinfo {volume} {120}},\ \bibinfo {pages}
  {7976--7986} (\bibinfo {year} {2016})}\BibitemShut {NoStop}%
\bibitem [{\citenamefont {Carignano}\ \emph {et~al.}(2015)\citenamefont
  {Carignano}, \citenamefont {Kachmar},\ and\ \citenamefont
  {Hutter}}]{Carignano2015}%
  \BibitemOpen
  \bibfield  {author} {\bibinfo {author} {\bibfnamefont {Marcelo~A.}\
  \bibnamefont {Carignano}}, \bibinfo {author} {\bibfnamefont {Ali}\
  \bibnamefont {Kachmar}}, \ and\ \bibinfo {author} {\bibfnamefont
  {J{\"{u}}rg}\ \bibnamefont {Hutter}},\ }\bibfield  {title} {\enquote
  {\bibinfo {title} {{Thermal Effects on CH 3 NH 3 PbI 3 Perovskite from Ab
  Initio Molecular Dynamics Simulations}},}\ }\href {\doibase
  10.1021/jp510568n} {\bibfield  {journal} {\bibinfo  {journal} {J. Phys. Chem.
  C}\ }\textbf {\bibinfo {volume} {119}},\ \bibinfo {pages} {8991--8997}
  (\bibinfo {year} {2015})}\BibitemShut {NoStop}%
\bibitem [{\citenamefont {Mattoni}\ \emph {et~al.}(2015)\citenamefont
  {Mattoni}, \citenamefont {Filippetti}, \citenamefont {Saba},\ and\
  \citenamefont {Delugas}}]{Mattoni2015a}%
  \BibitemOpen
  \bibfield  {author} {\bibinfo {author} {\bibfnamefont {A.}~\bibnamefont
  {Mattoni}}, \bibinfo {author} {\bibfnamefont {A.}~\bibnamefont {Filippetti}},
  \bibinfo {author} {\bibfnamefont {M.~I.}\ \bibnamefont {Saba}}, \ and\
  \bibinfo {author} {\bibfnamefont {P.}~\bibnamefont {Delugas}},\ }\bibfield
  {title} {\enquote {\bibinfo {title} {{Methylammonium Rotational Dynamics in
  Lead Halide Perovskite by Classical Molecular Dynamics: The Role of
  Temperature}},}\ }\href {\doibase 10.1021/acs.jpcc.5b04283} {\bibfield
  {journal} {\bibinfo  {journal} {J. Phys. Chem. C}\ }\textbf {\bibinfo
  {volume} {119}},\ \bibinfo {pages} {17421--17428} (\bibinfo {year}
  {2015})}\BibitemShut {NoStop}%
\bibitem [{\citenamefont {Quarti}\ \emph {et~al.}(2014)\citenamefont {Quarti},
  \citenamefont {Mosconi},\ and\ \citenamefont {{De Angelis}}}]{Quarti2014}%
  \BibitemOpen
  \bibfield  {author} {\bibinfo {author} {\bibfnamefont {Claudio}\ \bibnamefont
  {Quarti}}, \bibinfo {author} {\bibfnamefont {Edoardo}\ \bibnamefont
  {Mosconi}}, \ and\ \bibinfo {author} {\bibfnamefont {Filippo}\ \bibnamefont
  {{De Angelis}}},\ }\bibfield  {title} {\enquote {\bibinfo {title} {{Interplay
  of orientational order and electronic structure in methylammonium lead
  iodide: Implications for solar cell operation}},}\ }\href {\doibase
  10.1021/cm5032046} {\bibfield  {journal} {\bibinfo  {journal} {Chem. Mater.}\
  }\textbf {\bibinfo {volume} {26}},\ \bibinfo {pages} {6557--6569} (\bibinfo
  {year} {2014})}\BibitemShut {NoStop}%
\bibitem [{\citenamefont {Quarti}\ \emph {et~al.}(2015)\citenamefont {Quarti},
  \citenamefont {Mosconi},\ and\ \citenamefont {{De Angelis}}}]{Quarti2015}%
  \BibitemOpen
  \bibfield  {author} {\bibinfo {author} {\bibfnamefont {Claudio}\ \bibnamefont
  {Quarti}}, \bibinfo {author} {\bibfnamefont {Edoardo}\ \bibnamefont
  {Mosconi}}, \ and\ \bibinfo {author} {\bibfnamefont {Filippo}\ \bibnamefont
  {{De Angelis}}},\ }\bibfield  {title} {\enquote {\bibinfo {title}
  {{Structural and electronic properties of organo-halide hybrid perovskites
  from ab initio molecular dynamics}},}\ }\href {\doibase 10.1039/C5CP00599J}
  {\bibfield  {journal} {\bibinfo  {journal} {Phys. Chem. Chem. Phys.}\
  }\textbf {\bibinfo {volume} {17}},\ \bibinfo {pages} {9394--9409} (\bibinfo
  {year} {2015})}\BibitemShut {NoStop}%
\bibitem [{\citenamefont {Zhang}\ \emph {et~al.}(2018)\citenamefont {Zhang},
  \citenamefont {Geng}, \citenamefont {Tong}, \citenamefont {Chen},
  \citenamefont {Cao},\ and\ \citenamefont {Chen}}]{Zhang2018}%
  \BibitemOpen
  \bibfield  {author} {\bibinfo {author} {\bibfnamefont {Le}~\bibnamefont
  {Zhang}}, \bibinfo {author} {\bibfnamefont {Wei}\ \bibnamefont {Geng}},
  \bibinfo {author} {\bibfnamefont {Chuan~Jia}\ \bibnamefont {Tong}}, \bibinfo
  {author} {\bibfnamefont {Xueguang}\ \bibnamefont {Chen}}, \bibinfo {author}
  {\bibfnamefont {Tengfei}\ \bibnamefont {Cao}}, \ and\ \bibinfo {author}
  {\bibfnamefont {Mingyang}\ \bibnamefont {Chen}},\ }\bibfield  {title}
  {\enquote {\bibinfo {title} {{Strain induced electronic structure variation
  in methyl-ammonium lead iodide perovskite}},}\ }\href {\doibase
  10.1038/s41598-018-25772-3} {\bibfield  {journal} {\bibinfo  {journal} {Sci.
  Rep.}\ }\textbf {\bibinfo {volume} {8}},\ \bibinfo {pages} {7760} (\bibinfo
  {year} {2018})}\BibitemShut {NoStop}%
\bibitem [{\citenamefont {Senn}\ \emph {et~al.}(2016)\citenamefont {Senn},
  \citenamefont {Keen}, \citenamefont {Lucas}, \citenamefont {Hriljac},\ and\
  \citenamefont {Goodwin}}]{Senn2016}%
  \BibitemOpen
  \bibfield  {author} {\bibinfo {author} {\bibfnamefont {M.~S.}\ \bibnamefont
  {Senn}}, \bibinfo {author} {\bibfnamefont {D.~A.}\ \bibnamefont {Keen}},
  \bibinfo {author} {\bibfnamefont {T.~C.A.}\ \bibnamefont {Lucas}}, \bibinfo
  {author} {\bibfnamefont {J.~A.}\ \bibnamefont {Hriljac}}, \ and\ \bibinfo
  {author} {\bibfnamefont {A.~L.}\ \bibnamefont {Goodwin}},\ }\bibfield
  {title} {\enquote {\bibinfo {title} {{Emergence of Long-Range Order in BaTiO3
  from Local Symmetry-Breaking Distortions}},}\ }\href {\doibase
  10.1103/PhysRevLett.116.207602} {\bibfield  {journal} {\bibinfo  {journal}
  {Phys. Rev. Lett.}\ }\textbf {\bibinfo {volume} {116}},\ \bibinfo {pages}
  {207602} (\bibinfo {year} {2016})}\BibitemShut {NoStop}%
\bibitem [{\citenamefont {Bird}\ \emph {et~al.}(2020)\citenamefont {Bird},
  \citenamefont {Woodland-Scott}, \citenamefont {Hu}, \citenamefont {Wharmby},
  \citenamefont {Chen}, \citenamefont {Goodwin},\ and\ \citenamefont
  {Senn}}]{Bird2020}%
  \BibitemOpen
  \bibfield  {author} {\bibinfo {author} {\bibfnamefont {T.~A.}\ \bibnamefont
  {Bird}}, \bibinfo {author} {\bibfnamefont {J.}~\bibnamefont
  {Woodland-Scott}}, \bibinfo {author} {\bibfnamefont {L.}~\bibnamefont {Hu}},
  \bibinfo {author} {\bibfnamefont {M.~T.}\ \bibnamefont {Wharmby}}, \bibinfo
  {author} {\bibfnamefont {J.}~\bibnamefont {Chen}}, \bibinfo {author}
  {\bibfnamefont {A.~L.}\ \bibnamefont {Goodwin}}, \ and\ \bibinfo {author}
  {\bibfnamefont {M.~S.}\ \bibnamefont {Senn}},\ }\bibfield  {title} {\enquote
  {\bibinfo {title} {{Anharmonicity and scissoring modes in the negative
  thermal expansion materials ScF 3 and CaZrF 6}},}\ }\href {\doibase
  10.1103/physrevb.101.064306} {\bibfield  {journal} {\bibinfo  {journal}
  {Phys. Rev. B}\ }\textbf {\bibinfo {volume} {101}},\ \bibinfo {pages}
  {064306} (\bibinfo {year} {2020})}\BibitemShut {NoStop}%
\bibitem [{\citenamefont {Saidaminov}\ \emph {et~al.}(2015)\citenamefont
  {Saidaminov}, \citenamefont {Abdelhady}, \citenamefont {Murali},
  \citenamefont {Alarousu}, \citenamefont {Burlakov}, \citenamefont {Peng},
  \citenamefont {Dursun}, \citenamefont {Wang}, \citenamefont {He},
  \citenamefont {MacUlan}, \citenamefont {Goriely}, \citenamefont {Wu},
  \citenamefont {Mohammed},\ and\ \citenamefont {Bakr}}]{Saidaminov2015}%
  \BibitemOpen
  \bibfield  {author} {\bibinfo {author} {\bibfnamefont {Makhsud~I.}\
  \bibnamefont {Saidaminov}}, \bibinfo {author} {\bibfnamefont {Ahmed~L.}\
  \bibnamefont {Abdelhady}}, \bibinfo {author} {\bibfnamefont {Banavoth}\
  \bibnamefont {Murali}}, \bibinfo {author} {\bibfnamefont {Erkki}\
  \bibnamefont {Alarousu}}, \bibinfo {author} {\bibfnamefont {Victor~M.}\
  \bibnamefont {Burlakov}}, \bibinfo {author} {\bibfnamefont {Wei}\
  \bibnamefont {Peng}}, \bibinfo {author} {\bibfnamefont {Ibrahim}\
  \bibnamefont {Dursun}}, \bibinfo {author} {\bibfnamefont {Lingfei}\
  \bibnamefont {Wang}}, \bibinfo {author} {\bibfnamefont {Yao}\ \bibnamefont
  {He}}, \bibinfo {author} {\bibfnamefont {Giacomo}\ \bibnamefont {MacUlan}},
  \bibinfo {author} {\bibfnamefont {Alain}\ \bibnamefont {Goriely}}, \bibinfo
  {author} {\bibfnamefont {Tom}\ \bibnamefont {Wu}}, \bibinfo {author}
  {\bibfnamefont {Omar~F.}\ \bibnamefont {Mohammed}}, \ and\ \bibinfo {author}
  {\bibfnamefont {Osman~M.}\ \bibnamefont {Bakr}},\ }\bibfield  {title}
  {\enquote {\bibinfo {title} {{High-quality bulk hybrid perovskite single
  crystals within minutes by inverse temperature crystallization}},}\ }\href
  {\doibase 10.1038/ncomms8586} {\bibfield  {journal} {\bibinfo  {journal}
  {Nat. Commun.}\ }\textbf {\bibinfo {volume} {6}},\ \bibinfo {pages} {7586}
  (\bibinfo {year} {2015})}\BibitemShut {NoStop}%
\bibitem [{\citenamefont {Dippel}\ \emph {et~al.}(2015)\citenamefont {Dippel},
  \citenamefont {Liermann}, \citenamefont {Delitz}, \citenamefont {Walter},
  \citenamefont {Schulte-Schrepping}, \citenamefont {Seeck},\ and\
  \citenamefont {Franz}}]{Dippel2015}%
  \BibitemOpen
  \bibfield  {author} {\bibinfo {author} {\bibfnamefont {Ann-Christin}\
  \bibnamefont {Dippel}}, \bibinfo {author} {\bibfnamefont {Hanns-Peter}\
  \bibnamefont {Liermann}}, \bibinfo {author} {\bibfnamefont {Jan~Torben}\
  \bibnamefont {Delitz}}, \bibinfo {author} {\bibfnamefont {Peter}\
  \bibnamefont {Walter}}, \bibinfo {author} {\bibfnamefont {Horst}\
  \bibnamefont {Schulte-Schrepping}}, \bibinfo {author} {\bibfnamefont
  {Oliver~H.}\ \bibnamefont {Seeck}}, \ and\ \bibinfo {author} {\bibfnamefont
  {Hermann}\ \bibnamefont {Franz}},\ }\bibfield  {title} {\enquote {\bibinfo
  {title} {{Beamline P02.1 at PETRA III for high-resolution and high-energy
  powder diffraction}},}\ }\href {\doibase 10.1107/s1600577515002222}
  {\bibfield  {journal} {\bibinfo  {journal} {J. Synchrotron Radiat.}\ }\textbf
  {\bibinfo {volume} {22}},\ \bibinfo {pages} {675--687} (\bibinfo {year}
  {2015})}\BibitemShut {NoStop}%
\bibitem [{\citenamefont {Basham}\ \emph {et~al.}(2015)\citenamefont {Basham},
  \citenamefont {Filik}, \citenamefont {Wharmby}, \citenamefont {Chang},
  \citenamefont {{El Kassaby}}, \citenamefont {Gerring}, \citenamefont
  {Aishima}, \citenamefont {Levik}, \citenamefont {Pulford}, \citenamefont
  {Sikharulidze}, \citenamefont {Sneddon}, \citenamefont {Webber},
  \citenamefont {Dhesi}, \citenamefont {Maccherozzi}, \citenamefont {Svensson},
  \citenamefont {Brockhauser}, \citenamefont {N{\'{a}}ray},\ and\ \citenamefont
  {Ashton}}]{Basham2015}%
  \BibitemOpen
  \bibfield  {author} {\bibinfo {author} {\bibfnamefont {Mark}\ \bibnamefont
  {Basham}}, \bibinfo {author} {\bibfnamefont {Jacob}\ \bibnamefont {Filik}},
  \bibinfo {author} {\bibfnamefont {Michael~T.}\ \bibnamefont {Wharmby}},
  \bibinfo {author} {\bibfnamefont {Peter~C.Y.}\ \bibnamefont {Chang}},
  \bibinfo {author} {\bibfnamefont {Baha}\ \bibnamefont {{El Kassaby}}},
  \bibinfo {author} {\bibfnamefont {Matthew}\ \bibnamefont {Gerring}}, \bibinfo
  {author} {\bibfnamefont {Jun}\ \bibnamefont {Aishima}}, \bibinfo {author}
  {\bibfnamefont {Karl}\ \bibnamefont {Levik}}, \bibinfo {author}
  {\bibfnamefont {Bill~C.A.}\ \bibnamefont {Pulford}}, \bibinfo {author}
  {\bibfnamefont {Irakli}\ \bibnamefont {Sikharulidze}}, \bibinfo {author}
  {\bibfnamefont {Duncan}\ \bibnamefont {Sneddon}}, \bibinfo {author}
  {\bibfnamefont {Matthew}\ \bibnamefont {Webber}}, \bibinfo {author}
  {\bibfnamefont {Sarnjeet~S.}\ \bibnamefont {Dhesi}}, \bibinfo {author}
  {\bibfnamefont {Francesco}\ \bibnamefont {Maccherozzi}}, \bibinfo {author}
  {\bibfnamefont {Olof}\ \bibnamefont {Svensson}}, \bibinfo {author}
  {\bibfnamefont {Sandor}\ \bibnamefont {Brockhauser}}, \bibinfo {author}
  {\bibfnamefont {Gabor}\ \bibnamefont {N{\'{a}}ray}}, \ and\ \bibinfo {author}
  {\bibfnamefont {Alun~W.}\ \bibnamefont {Ashton}},\ }\bibfield  {title}
  {\enquote {\bibinfo {title} {{Data Analysis WorkbeNch (DAWN)}},}\ }\href
  {\doibase 10.1107/S1600577515002283} {\bibfield  {journal} {\bibinfo
  {journal} {J. Synchrotron Radiat.}\ }\textbf {\bibinfo {volume} {22}},\
  \bibinfo {pages} {853--858} (\bibinfo {year} {2015})}\BibitemShut {NoStop}%
\bibitem [{\citenamefont {McLain}\ \emph {et~al.}(2012)\citenamefont {McLain},
  \citenamefont {Bowron}, \citenamefont {Hannon},\ and\ \citenamefont
  {Soper}}]{McLain2012}%
  \BibitemOpen
  \bibfield  {author} {\bibinfo {author} {\bibfnamefont {S.~E.}\ \bibnamefont
  {McLain}}, \bibinfo {author} {\bibfnamefont {D.~T.}\ \bibnamefont {Bowron}},
  \bibinfo {author} {\bibfnamefont {A.~C.}\ \bibnamefont {Hannon}}, \ and\
  \bibinfo {author} {\bibfnamefont {A.~K.}\ \bibnamefont {Soper}},\ }\bibfield
  {title} {\enquote {\bibinfo {title} {{GUDRUN, a computer program developed
  for analysis of neutron diffraction data, Chilton: ISIS Facility, Rutherford
  Appleton Laboratory}},}\ }\href@noop {} {\  (\bibinfo {year}
  {2012})}\BibitemShut {NoStop}%
\bibitem [{\citenamefont {Campbell}\ \emph {et~al.}(2006)\citenamefont
  {Campbell}, \citenamefont {Stokes}, \citenamefont {Tanner},\ and\
  \citenamefont {Hatch}}]{Campbell2006}%
  \BibitemOpen
  \bibfield  {author} {\bibinfo {author} {\bibfnamefont {Branton~J}\
  \bibnamefont {Campbell}}, \bibinfo {author} {\bibfnamefont {Harold~T}\
  \bibnamefont {Stokes}}, \bibinfo {author} {\bibfnamefont {David~E}\
  \bibnamefont {Tanner}}, \ and\ \bibinfo {author} {\bibfnamefont {Dorian~M}\
  \bibnamefont {Hatch}},\ }\bibfield  {title} {\enquote {\bibinfo {title}
  {{ISODISPLACE: A web-based tool for exploring structural distortions}},}\
  }\href {\doibase 10.1107/S0021889806014075} {\bibfield  {journal} {\bibinfo
  {journal} {J. Appl. Crystallogr.}\ }\textbf {\bibinfo {volume} {39}},\
  \bibinfo {pages} {607--614} (\bibinfo {year} {2006})}\BibitemShut {NoStop}%
\bibitem [{\citenamefont {Evans}(2010)}]{Evans2010}%
  \BibitemOpen
  \bibfield  {author} {\bibinfo {author} {\bibfnamefont {John~S.O.}\
  \bibnamefont {Evans}},\ }\bibfield  {title} {\enquote {\bibinfo {title}
  {{Advanced input files \& parametric quantitative analysis using topas}},}\
  }\href {\doibase 10.4028/www.scientific.net/MSF.651.1} {\bibfield  {journal}
  {\bibinfo  {journal} {Mater. Sci. Forum}\ }\textbf {\bibinfo {volume}
  {651}},\ \bibinfo {pages} {1--9} (\bibinfo {year} {2010})}\BibitemShut
  {NoStop}%
\bibitem [{\citenamefont {Page}\ \emph {et~al.}(2016)\citenamefont {Page},
  \citenamefont {Siewenie}, \citenamefont {Quadrelli},\ and\ \citenamefont
  {Malavasi}}]{Page2016}%
  \BibitemOpen
  \bibfield  {author} {\bibinfo {author} {\bibfnamefont {Katharine}\
  \bibnamefont {Page}}, \bibinfo {author} {\bibfnamefont {Joan~E.}\
  \bibnamefont {Siewenie}}, \bibinfo {author} {\bibfnamefont {Paolo}\
  \bibnamefont {Quadrelli}}, \ and\ \bibinfo {author} {\bibfnamefont {Lorenzo}\
  \bibnamefont {Malavasi}},\ }\bibfield  {title} {\enquote {\bibinfo {title}
  {{Short-Range Order of Methylammonium and Persistence of Distortion at the
  Local Scale in MAPbBr3Hybrid Perovskite}},}\ }\href {\doibase
  10.1002/anie.201608602} {\bibfield  {journal} {\bibinfo  {journal} {Angew.
  Chem. Int. Ed.}\ }\textbf {\bibinfo {volume} {55}},\ \bibinfo {pages}
  {14320--14324} (\bibinfo {year} {2016})}\BibitemShut {NoStop}%
\bibitem [{\citenamefont {Bernasconi}\ and\ \citenamefont
  {Malavasi}(2017)}]{Bernasconi2017}%
  \BibitemOpen
  \bibfield  {author} {\bibinfo {author} {\bibfnamefont {Andrea}\ \bibnamefont
  {Bernasconi}}\ and\ \bibinfo {author} {\bibfnamefont {Lorenzo}\ \bibnamefont
  {Malavasi}},\ }\bibfield  {title} {\enquote {\bibinfo {title} {{Direct
  evidence of permanent octahedra distortion in MAPbBr3 hybrid perovskite}},}\
  }\href {\doibase 10.1021/acsenergylett.7b00139} {\bibfield  {journal}
  {\bibinfo  {journal} {ACS Energy Lett.}\ }\textbf {\bibinfo {volume} {2}},\
  \bibinfo {pages} {863--868} (\bibinfo {year} {2017})}\BibitemShut {NoStop}%
\bibitem [{\citenamefont {Bernasconi}\ \emph {et~al.}(2018)\citenamefont
  {Bernasconi}, \citenamefont {Page}, \citenamefont {Dai}, \citenamefont {Tan},
  \citenamefont {Rappe},\ and\ \citenamefont {Malavasi}}]{Bernasconi2018}%
  \BibitemOpen
  \bibfield  {author} {\bibinfo {author} {\bibfnamefont {Andrea}\ \bibnamefont
  {Bernasconi}}, \bibinfo {author} {\bibfnamefont {Katharine}\ \bibnamefont
  {Page}}, \bibinfo {author} {\bibfnamefont {Zhenbang}\ \bibnamefont {Dai}},
  \bibinfo {author} {\bibfnamefont {Liang~Z.}\ \bibnamefont {Tan}}, \bibinfo
  {author} {\bibfnamefont {Andrew~M.}\ \bibnamefont {Rappe}}, \ and\ \bibinfo
  {author} {\bibfnamefont {Lorenzo}\ \bibnamefont {Malavasi}},\ }\bibfield
  {title} {\enquote {\bibinfo {title} {{Ubiquitous Short-Range Distortion of
  Hybrid Perovskites and Hydrogen-Bonding Role: The MAPbCl 3 Case}},}\ }\href
  {\doibase 10.1021/acs.jpcc.8b10086} {\bibfield  {journal} {\bibinfo
  {journal} {J. Phys. Chem. C}\ }\textbf {\bibinfo {volume} {122}},\ \bibinfo
  {pages} {28265--28272} (\bibinfo {year} {2018})}\BibitemShut {NoStop}%
\bibitem [{\citenamefont {Kresse}\ and\ \citenamefont
  {Hafner}(1994)}]{Kresse1994}%
  \BibitemOpen
  \bibfield  {author} {\bibinfo {author} {\bibfnamefont {G.}~\bibnamefont
  {Kresse}}\ and\ \bibinfo {author} {\bibfnamefont {J.}~\bibnamefont
  {Hafner}},\ }\bibfield  {title} {\enquote {\bibinfo {title} {{Ab initio
  molecular-dynamics simulation of the liquid-metalamorphous- semiconductor
  transition in germanium}},}\ }\href {\doibase 10.1103/PhysRevB.49.14251}
  {\bibfield  {journal} {\bibinfo  {journal} {Phys. Rev. B}\ }\textbf {\bibinfo
  {volume} {49}},\ \bibinfo {pages} {14251--14269} (\bibinfo {year}
  {1994})}\BibitemShut {NoStop}%
\bibitem [{\citenamefont {Kresse}\ and\ \citenamefont
  {Furthm{\"{u}}ller}(1996{\natexlab{a}})}]{Kresse1996}%
  \BibitemOpen
  \bibfield  {author} {\bibinfo {author} {\bibfnamefont {G.}~\bibnamefont
  {Kresse}}\ and\ \bibinfo {author} {\bibfnamefont {J.}~\bibnamefont
  {Furthm{\"{u}}ller}},\ }\bibfield  {title} {\enquote {\bibinfo {title}
  {{Efficient iterative schemes for ab initio total-energy calculations using a
  plane-wave basis set}},}\ }\href {\doibase 10.1021/acs.jpca.0c01375}
  {\bibfield  {journal} {\bibinfo  {journal} {Phys. Rev. B}\ }\textbf {\bibinfo
  {volume} {54}},\ \bibinfo {pages} {11169 -- 11186} (\bibinfo {year}
  {1996}{\natexlab{a}})}\BibitemShut {NoStop}%
\bibitem [{\citenamefont {Kresse}\ and\ \citenamefont
  {Furthm{\"{u}}ller}(1996{\natexlab{b}})}]{Kresse1996b}%
  \BibitemOpen
  \bibfield  {author} {\bibinfo {author} {\bibfnamefont {G.}~\bibnamefont
  {Kresse}}\ and\ \bibinfo {author} {\bibfnamefont {J.}~\bibnamefont
  {Furthm{\"{u}}ller}},\ }\bibfield  {title} {\enquote {\bibinfo {title}
  {{Efficiency of ab-initio total energy calculations for metals and
  semiconductors using a plane-wave basis set}},}\ }\href {\doibase
  10.1016/0927-0256(96)00008-0} {\bibfield  {journal} {\bibinfo  {journal}
  {Comput. Mater. Sci.}\ }\textbf {\bibinfo {volume} {6}},\ \bibinfo {pages}
  {15--50} (\bibinfo {year} {1996}{\natexlab{b}})}\BibitemShut {NoStop}%
\bibitem [{\citenamefont {Kresse}\ and\ \citenamefont
  {Hafner}(1993)}]{Kresse1993}%
  \BibitemOpen
  \bibfield  {author} {\bibinfo {author} {\bibfnamefont {G.}~\bibnamefont
  {Kresse}}\ and\ \bibinfo {author} {\bibfnamefont {J.}~\bibnamefont
  {Hafner}},\ }\bibfield  {title} {\enquote {\bibinfo {title} {{Ab initio
  molecular dynamics for liquid metals}},}\ }\href {\doibase
  10.1016/0022-3093(95)00355-X} {\bibfield  {journal} {\bibinfo  {journal} {J.
  Non-Cryst. Solids}\ }\textbf {\bibinfo {volume} {47}},\ \bibinfo {pages}
  {558--561} (\bibinfo {year} {1993})}\BibitemShut {NoStop}%
\bibitem [{\citenamefont {Klimes}\ \emph {et~al.}(2011)\citenamefont {Klimes},
  \citenamefont {Bowler},\ and\ \citenamefont {Michaelides}}]{Klime2011}%
  \BibitemOpen
  \bibfield  {author} {\bibinfo {author} {\bibfnamefont {Jiř{\'{i}}}\
  \bibnamefont {Klimes}}, \bibinfo {author} {\bibfnamefont {David~R.}\
  \bibnamefont {Bowler}}, \ and\ \bibinfo {author} {\bibfnamefont {Angelos}\
  \bibnamefont {Michaelides}},\ }\bibfield  {title} {\enquote {\bibinfo {title}
  {{Van der Waals density functionals applied to solids}},}\ }\href {\doibase
  10.1103/PhysRevB.83.195131} {\bibfield  {journal} {\bibinfo  {journal} {Phys.
  Rev. B}\ }\textbf {\bibinfo {volume} {83}},\ \bibinfo {pages} {195131}
  (\bibinfo {year} {2011})}\BibitemShut {NoStop}%
\bibitem [{\citenamefont {Blochl}(1994)}]{Blochl1994}%
  \BibitemOpen
  \bibfield  {author} {\bibinfo {author} {\bibfnamefont {P~E}\ \bibnamefont
  {Blochl}},\ }\bibfield  {title} {\enquote {\bibinfo {title} {{Projector
  augmented-wave method}},}\ }\href@noop {} {\bibfield  {journal} {\bibinfo
  {journal} {Phys. Rev. B}\ }\textbf {\bibinfo {volume} {50}},\ \bibinfo
  {pages} {17953--17979} (\bibinfo {year} {1994})}\BibitemShut {NoStop}%
\bibitem [{\citenamefont {Beecher}\ \emph {et~al.}(2016)\citenamefont
  {Beecher}, \citenamefont {Semonin}, \citenamefont {Skelton}, \citenamefont
  {Frost}, \citenamefont {Terban}, \citenamefont {Zhai}, \citenamefont
  {Alatas}, \citenamefont {Owen}, \citenamefont {Walsh},\ and\ \citenamefont
  {Billinge}}]{Beecher2016}%
  \BibitemOpen
  \bibfield  {author} {\bibinfo {author} {\bibfnamefont {Alexander~N.}\
  \bibnamefont {Beecher}}, \bibinfo {author} {\bibfnamefont {Octavi~E.}\
  \bibnamefont {Semonin}}, \bibinfo {author} {\bibfnamefont {Jonathan~M.}\
  \bibnamefont {Skelton}}, \bibinfo {author} {\bibfnamefont {Jarvist~M.}\
  \bibnamefont {Frost}}, \bibinfo {author} {\bibfnamefont {Maxwell~W.}\
  \bibnamefont {Terban}}, \bibinfo {author} {\bibfnamefont {Haowei}\
  \bibnamefont {Zhai}}, \bibinfo {author} {\bibfnamefont {Ahmet}\ \bibnamefont
  {Alatas}}, \bibinfo {author} {\bibfnamefont {Jonathan~S.}\ \bibnamefont
  {Owen}}, \bibinfo {author} {\bibfnamefont {Aron}\ \bibnamefont {Walsh}}, \
  and\ \bibinfo {author} {\bibfnamefont {Simon~J.L.}\ \bibnamefont
  {Billinge}},\ }\bibfield  {title} {\enquote {\bibinfo {title} {{Direct
  Observation of Dynamic Symmetry Breaking above Room Temperature in
  Methylammonium Lead Iodide Perovskite}},}\ }\href {\doibase
  10.1021/acsenergylett.6b00381} {\bibfield  {journal} {\bibinfo  {journal}
  {ACS Energy Lett.}\ }\textbf {\bibinfo {volume} {1}},\ \bibinfo {pages}
  {880--887} (\bibinfo {year} {2016})}\BibitemShut {NoStop}%
\bibitem [{\citenamefont {Liu}(2017)}]{Liu2017}%
  \BibitemOpen
  \bibfield  {author} {\bibinfo {author} {\bibfnamefont {Jiaxun}\ \bibnamefont
  {Liu}},\ }\emph {\bibinfo {title} {{Local structure of lead halide
  perovskites for photovoltaic applications}}},\ \href {\doibase
  10.1107/s2053273317084571} {\bibinfo {type} {Doctor of philosophy}},\
  \bibinfo  {school} {Queen Mary, University of London} (\bibinfo {year}
  {2017})\BibitemShut {NoStop}%
\bibitem [{\citenamefont {Mosconi}\ \emph {et~al.}(2016)\citenamefont
  {Mosconi}, \citenamefont {Umari},\ and\ \citenamefont {{De
  Angelis}}}]{Mosconi2016}%
  \BibitemOpen
  \bibfield  {author} {\bibinfo {author} {\bibfnamefont {Edoardo}\ \bibnamefont
  {Mosconi}}, \bibinfo {author} {\bibfnamefont {Paolo}\ \bibnamefont {Umari}},
  \ and\ \bibinfo {author} {\bibfnamefont {Filippo}\ \bibnamefont {{De
  Angelis}}},\ }\bibfield  {title} {\enquote {\bibinfo {title} {{Electronic and
  optical properties of MAPbX3 perovskites (X = I, Br, Cl): A unified DFT and
  GW theoretical analysis}},}\ }\href {\doibase 10.1039/c6cp03969c} {\bibfield
  {journal} {\bibinfo  {journal} {Phys. Chem. Chem. Phys.}\ }\textbf {\bibinfo
  {volume} {18}},\ \bibinfo {pages} {27158--27164} (\bibinfo {year}
  {2016})}\BibitemShut {NoStop}%
\bibitem [{\citenamefont {Papavassiliou}\ and\ \citenamefont
  {Koutselas}(1995)}]{Papavassiliou1995}%
  \BibitemOpen
  \bibfield  {author} {\bibinfo {author} {\bibfnamefont {G.~C.}\ \bibnamefont
  {Papavassiliou}}\ and\ \bibinfo {author} {\bibfnamefont {I.~B.}\ \bibnamefont
  {Koutselas}},\ }\bibfield  {title} {\enquote {\bibinfo {title} {{Structural,
  optical and related properties of some natural three- and lower-dimensional
  semiconductor systems}},}\ }\href {\doibase 10.1016/0379-6779(94)03017-Z}
  {\bibfield  {journal} {\bibinfo  {journal} {Synth. Met.}\ }\textbf {\bibinfo
  {volume} {71}},\ \bibinfo {pages} {1713--1714} (\bibinfo {year}
  {1995})}\BibitemShut {NoStop}%
\bibitem [{\citenamefont {Frost}(2017)}]{Frost2017}%
  \BibitemOpen
  \bibfield  {author} {\bibinfo {author} {\bibfnamefont {Jarvist~Moore}\
  \bibnamefont {Frost}},\ }\bibfield  {title} {\enquote {\bibinfo {title}
  {{Calculating polaron mobility in halide perovskites}},}\ }\href {\doibase
  10.1103/PhysRevB.96.195202} {\bibfield  {journal} {\bibinfo  {journal} {Phys.
  Rev. B}\ }\textbf {\bibinfo {volume} {96}},\ \bibinfo {pages} {195202}
  (\bibinfo {year} {2017})}\BibitemShut {NoStop}%
\bibitem [{\citenamefont {Bonn}\ \emph {et~al.}(2017)\citenamefont {Bonn},
  \citenamefont {Miyata}, \citenamefont {Hendry},\ and\ \citenamefont
  {Zhu}}]{Bonn2017}%
  \BibitemOpen
  \bibfield  {author} {\bibinfo {author} {\bibfnamefont {Mischa}\ \bibnamefont
  {Bonn}}, \bibinfo {author} {\bibfnamefont {Kiyoshi}\ \bibnamefont {Miyata}},
  \bibinfo {author} {\bibfnamefont {Euan}\ \bibnamefont {Hendry}}, \ and\
  \bibinfo {author} {\bibfnamefont {X.~Y.}\ \bibnamefont {Zhu}},\ }\bibfield
  {title} {\enquote {\bibinfo {title} {{Role of Dielectric Drag in Polaron
  Mobility in Lead Halide Perovskites}},}\ }\href {\doibase
  10.1021/acsenergylett.7b00717} {\bibfield  {journal} {\bibinfo  {journal}
  {ACS Energy Lett.}\ }\textbf {\bibinfo {volume} {2}},\ \bibinfo {pages}
  {2555--2562} (\bibinfo {year} {2017})}\BibitemShut {NoStop}%
\bibitem [{\citenamefont {Brivio}\ \emph {et~al.}(2014)\citenamefont {Brivio},
  \citenamefont {Butler}, \citenamefont {Walsh},\ and\ \citenamefont {{Van
  Schilfgaarde}}}]{Brivio2014}%
  \BibitemOpen
  \bibfield  {author} {\bibinfo {author} {\bibfnamefont {Federico}\
  \bibnamefont {Brivio}}, \bibinfo {author} {\bibfnamefont {Keith~T.}\
  \bibnamefont {Butler}}, \bibinfo {author} {\bibfnamefont {Aron}\ \bibnamefont
  {Walsh}}, \ and\ \bibinfo {author} {\bibfnamefont {Mark}\ \bibnamefont {{Van
  Schilfgaarde}}},\ }\bibfield  {title} {\enquote {\bibinfo {title}
  {{Relativistic quasiparticle self-consistent electronic structure of hybrid
  halide perovskite photovoltaic absorbers}},}\ }\href {\doibase
  10.1103/PhysRevB.89.155204} {\bibfield  {journal} {\bibinfo  {journal} {Phys.
  Rev. B}\ }\textbf {\bibinfo {volume} {89}},\ \bibinfo {pages} {155204}
  (\bibinfo {year} {2014})}\BibitemShut {NoStop}%
\bibitem [{\citenamefont {Even}\ \emph {et~al.}(2013)\citenamefont {Even},
  \citenamefont {Pedesseau}, \citenamefont {Jancu},\ and\ \citenamefont
  {Katan}}]{Even2013}%
  \BibitemOpen
  \bibfield  {author} {\bibinfo {author} {\bibfnamefont {Jacky}\ \bibnamefont
  {Even}}, \bibinfo {author} {\bibfnamefont {Laurent}\ \bibnamefont
  {Pedesseau}}, \bibinfo {author} {\bibfnamefont {Jean~Marc}\ \bibnamefont
  {Jancu}}, \ and\ \bibinfo {author} {\bibfnamefont {Claudine}\ \bibnamefont
  {Katan}},\ }\bibfield  {title} {\enquote {\bibinfo {title} {{Importance of
  spin-orbit coupling in hybrid organic/inorganic perovskites for photovoltaic
  applications}},}\ }\href {\doibase 10.1021/jz401532q} {\bibfield  {journal}
  {\bibinfo  {journal} {J. Phys. Chem. Lett.}\ }\textbf {\bibinfo {volume}
  {4}},\ \bibinfo {pages} {2999--3005} (\bibinfo {year} {2013})}\BibitemShut
  {NoStop}%
\bibitem [{\citenamefont {Umari}\ \emph {et~al.}(2014)\citenamefont {Umari},
  \citenamefont {Mosconi},\ and\ \citenamefont {{De Angelis}}}]{Umari2014}%
  \BibitemOpen
  \bibfield  {author} {\bibinfo {author} {\bibfnamefont {Paolo}\ \bibnamefont
  {Umari}}, \bibinfo {author} {\bibfnamefont {Edoardo}\ \bibnamefont
  {Mosconi}}, \ and\ \bibinfo {author} {\bibfnamefont {Filippo}\ \bibnamefont
  {{De Angelis}}},\ }\bibfield  {title} {\enquote {\bibinfo {title}
  {{Relativistic GW calculations on CH3 NH3 PbI 3 and CH3 NH3 SnI3 Perovskites
  for Solar Cell Applications}},}\ }\href {\doibase 10.1038/srep04467}
  {\bibfield  {journal} {\bibinfo  {journal} {Sci. Rep.}\ }\textbf {\bibinfo
  {volume} {4}},\ \bibinfo {pages} {4467} (\bibinfo {year} {2014})}\BibitemShut
  {NoStop}%
\bibitem [{\citenamefont {Li}\ \emph {et~al.}(2017)\citenamefont {Li},
  \citenamefont {Kawakita}, \citenamefont {Liu}, \citenamefont {Wang},
  \citenamefont {Matsuura}, \citenamefont {Shibata}, \citenamefont
  {Ohira-Kawamura}, \citenamefont {Yamada}, \citenamefont {Lin}, \citenamefont
  {Nakajima},\ and\ \citenamefont {Liu}}]{Li2017}%
  \BibitemOpen
  \bibfield  {author} {\bibinfo {author} {\bibfnamefont {Bing}\ \bibnamefont
  {Li}}, \bibinfo {author} {\bibfnamefont {Yukinobu}\ \bibnamefont {Kawakita}},
  \bibinfo {author} {\bibfnamefont {Yucheng}\ \bibnamefont {Liu}}, \bibinfo
  {author} {\bibfnamefont {Mingchao}\ \bibnamefont {Wang}}, \bibinfo {author}
  {\bibfnamefont {Masato}\ \bibnamefont {Matsuura}}, \bibinfo {author}
  {\bibfnamefont {Kaoru}\ \bibnamefont {Shibata}}, \bibinfo {author}
  {\bibfnamefont {Seiko}\ \bibnamefont {Ohira-Kawamura}}, \bibinfo {author}
  {\bibfnamefont {Takeshi}\ \bibnamefont {Yamada}}, \bibinfo {author}
  {\bibfnamefont {Shangchao}\ \bibnamefont {Lin}}, \bibinfo {author}
  {\bibfnamefont {Kenji}\ \bibnamefont {Nakajima}}, \ and\ \bibinfo {author}
  {\bibfnamefont {Shengzhong~Frank}\ \bibnamefont {Liu}},\ }\bibfield  {title}
  {\enquote {\bibinfo {title} {{Polar rotor scattering as atomic-level origin
  of low mobility and thermal conductivity of perovskite CH 3 NH 3 PbI 3}},}\
  }\href {\doibase 10.1038/ncomms16086} {\bibfield  {journal} {\bibinfo
  {journal} {Nat. Commun.}\ }\textbf {\bibinfo {volume} {8}},\ \bibinfo {pages}
  {16086} (\bibinfo {year} {2017})}\BibitemShut {NoStop}%
\bibitem [{\citenamefont {Ghosh}\ \emph
  {et~al.}(2017{\natexlab{b}})\citenamefont {Ghosh}, \citenamefont {Aharon},
  \citenamefont {Etgar},\ and\ \citenamefont {Ruhman}}]{Ghosh2017a}%
  \BibitemOpen
  \bibfield  {author} {\bibinfo {author} {\bibfnamefont {Tufan}\ \bibnamefont
  {Ghosh}}, \bibinfo {author} {\bibfnamefont {Sigalit}\ \bibnamefont {Aharon}},
  \bibinfo {author} {\bibfnamefont {Lioz}\ \bibnamefont {Etgar}}, \ and\
  \bibinfo {author} {\bibfnamefont {Sanford}\ \bibnamefont {Ruhman}},\
  }\bibfield  {title} {\enquote {\bibinfo {title} {{Free Carrier Emergence and
  Onset of Electron-Phonon Coupling in Methylammonium Lead Halide Perovskite
  Films}},}\ }\href {\doibase 10.1021/jacs.7b09508} {\bibfield  {journal}
  {\bibinfo  {journal} {J. Am. Chem. Soc.}\ }\textbf {\bibinfo {volume}
  {139}},\ \bibinfo {pages} {18262--18270} (\bibinfo {year}
  {2017}{\natexlab{b}})}\BibitemShut {NoStop}%
\bibitem [{\citenamefont {Yaffe}\ \emph {et~al.}(2017)\citenamefont {Yaffe},
  \citenamefont {Guo}, \citenamefont {Tan}, \citenamefont {Egger},
  \citenamefont {Hull}, \citenamefont {Stoumpos}, \citenamefont {Zheng},
  \citenamefont {Heinz}, \citenamefont {Kronik}, \citenamefont {Kanatzidis},
  \citenamefont {Owen}, \citenamefont {Rappe}, \citenamefont {Pimenta},\ and\
  \citenamefont {Brus}}]{Yaffe2017}%
  \BibitemOpen
  \bibfield  {author} {\bibinfo {author} {\bibfnamefont {Omer}\ \bibnamefont
  {Yaffe}}, \bibinfo {author} {\bibfnamefont {Yinsheng}\ \bibnamefont {Guo}},
  \bibinfo {author} {\bibfnamefont {Liang~Z.}\ \bibnamefont {Tan}}, \bibinfo
  {author} {\bibfnamefont {David~A.}\ \bibnamefont {Egger}}, \bibinfo {author}
  {\bibfnamefont {Trevor}\ \bibnamefont {Hull}}, \bibinfo {author}
  {\bibfnamefont {Constantinos~C.}\ \bibnamefont {Stoumpos}}, \bibinfo {author}
  {\bibfnamefont {Fan}\ \bibnamefont {Zheng}}, \bibinfo {author} {\bibfnamefont
  {Tony~F.}\ \bibnamefont {Heinz}}, \bibinfo {author} {\bibfnamefont {Leeor}\
  \bibnamefont {Kronik}}, \bibinfo {author} {\bibfnamefont {Mercouri~G.}\
  \bibnamefont {Kanatzidis}}, \bibinfo {author} {\bibfnamefont {Jonathan~S.}\
  \bibnamefont {Owen}}, \bibinfo {author} {\bibfnamefont {Andrew~M.}\
  \bibnamefont {Rappe}}, \bibinfo {author} {\bibfnamefont {Marcos~A.}\
  \bibnamefont {Pimenta}}, \ and\ \bibinfo {author} {\bibfnamefont {Louis~E.}\
  \bibnamefont {Brus}},\ }\bibfield  {title} {\enquote {\bibinfo {title}
  {{Local Polar Fluctuations in Lead Halide Perovskite Crystals}},}\ }\href
  {\doibase 10.1103/PhysRevLett.118.136001} {\bibfield  {journal} {\bibinfo
  {journal} {Phys. Rev. Lett.}\ }\textbf {\bibinfo {volume} {118}},\ \bibinfo
  {pages} {136001} (\bibinfo {year} {2017})}\BibitemShut {NoStop}%
\bibitem [{\citenamefont {Songvilay}\ \emph
  {et~al.}(2019{\natexlab{b}})\citenamefont {Songvilay}, \citenamefont
  {Giles-Donovan}, \citenamefont {Bari}, \citenamefont {Ye}, \citenamefont
  {Minns}, \citenamefont {Green}, \citenamefont {Xu}, \citenamefont {Gehring},
  \citenamefont {Schmalzl}, \citenamefont {Ratcliff}, \citenamefont {Brown},
  \citenamefont {Chernyshov}, \citenamefont {{Van Beek}}, \citenamefont
  {Cochran},\ and\ \citenamefont {Stock}}]{Songvilay2019a}%
  \BibitemOpen
  \bibfield  {author} {\bibinfo {author} {\bibfnamefont {M.}~\bibnamefont
  {Songvilay}}, \bibinfo {author} {\bibfnamefont {N.}~\bibnamefont
  {Giles-Donovan}}, \bibinfo {author} {\bibfnamefont {M.}~\bibnamefont {Bari}},
  \bibinfo {author} {\bibfnamefont {Z.~G.}\ \bibnamefont {Ye}}, \bibinfo
  {author} {\bibfnamefont {J.~L.}\ \bibnamefont {Minns}}, \bibinfo {author}
  {\bibfnamefont {M.~A.}\ \bibnamefont {Green}}, \bibinfo {author}
  {\bibfnamefont {Guangyong}\ \bibnamefont {Xu}}, \bibinfo {author}
  {\bibfnamefont {P.~M.}\ \bibnamefont {Gehring}}, \bibinfo {author}
  {\bibfnamefont {K.}~\bibnamefont {Schmalzl}}, \bibinfo {author}
  {\bibfnamefont {W.~D.}\ \bibnamefont {Ratcliff}}, \bibinfo {author}
  {\bibfnamefont {C.~M.}\ \bibnamefont {Brown}}, \bibinfo {author}
  {\bibfnamefont {D.}~\bibnamefont {Chernyshov}}, \bibinfo {author}
  {\bibfnamefont {W.}~\bibnamefont {{Van Beek}}}, \bibinfo {author}
  {\bibfnamefont {S.}~\bibnamefont {Cochran}}, \ and\ \bibinfo {author}
  {\bibfnamefont {C.}~\bibnamefont {Stock}},\ }\bibfield  {title} {\enquote
  {\bibinfo {title} {{Common acoustic phonon lifetimes in inorganic and hybrid
  lead halide perovskites}},}\ }\href {\doibase
  10.1103/PhysRevMaterials.3.093602} {\bibfield  {journal} {\bibinfo  {journal}
  {Phys. Rev. Mater.}\ }\textbf {\bibinfo {volume} {3}},\ \bibinfo {pages}
  {093602} (\bibinfo {year} {2019}{\natexlab{b}})}\BibitemShut {NoStop}%
\bibitem [{\citenamefont {Zhao}\ \emph {et~al.}(2020)\citenamefont {Zhao},
  \citenamefont {Dalpian}, \citenamefont {Wang},\ and\ \citenamefont
  {Zunger}}]{Zhao2020}%
  \BibitemOpen
  \bibfield  {author} {\bibinfo {author} {\bibfnamefont {Xin~Gang}\
  \bibnamefont {Zhao}}, \bibinfo {author} {\bibfnamefont {Gustavo~M.}\
  \bibnamefont {Dalpian}}, \bibinfo {author} {\bibfnamefont {Zhi}\ \bibnamefont
  {Wang}}, \ and\ \bibinfo {author} {\bibfnamefont {Alex}\ \bibnamefont
  {Zunger}},\ }\bibfield  {title} {\enquote {\bibinfo {title} {{Polymorphous
  nature of cubic halide perovskites}},}\ }\href {\doibase
  10.1103/PhysRevB.101.155137} {\bibfield  {journal} {\bibinfo  {journal}
  {Phys. Rev. B}\ }\textbf {\bibinfo {volume} {101}},\ \bibinfo {pages}
  {155137} (\bibinfo {year} {2020})}\BibitemShut {NoStop}%
\bibitem [{\citenamefont {Ganose}\ \emph {et~al.}(2018)\citenamefont {Ganose},
  \citenamefont {Jackson},\ and\ \citenamefont {Scanlon}}]{MGanose2018}%
  \BibitemOpen
  \bibfield  {author} {\bibinfo {author} {\bibfnamefont {Alex~M.}\ \bibnamefont
  {Ganose}}, \bibinfo {author} {\bibfnamefont {Adam~J.}\ \bibnamefont
  {Jackson}}, \ and\ \bibinfo {author} {\bibfnamefont {David~O.}\ \bibnamefont
  {Scanlon}},\ }\bibfield  {title} {\enquote {\bibinfo {title} {{sumo:
  Command-line tools for plotting and analysis of periodic ab initio
  calculations}},}\ }\href {\doibase 10.21105/joss.00717} {\bibfield  {journal}
  {\bibinfo  {journal} {J. Open Source Softw.}\ }\textbf {\bibinfo {volume}
  {3}},\ \bibinfo {pages} {717} (\bibinfo {year} {2018})}\BibitemShut {NoStop}%
\end{thebibliography}%
\end{document}